%% file: arxiv.tex
\documentclass[12pt]{article}
\usepackage[utf8]{inputenc}

\input{inserts/preamble.tex}

\usepackage[htt]{hyphenat}
\usepackage[%
    authordate,
    doi=false,
    isbn=false,
    eprint=false,
    giveninits=true,
    uniquename=mininit,
    maxcitenames=2,
    dashed=false,
    natbib=true,
    noibid
]{biblatex-chicago}
\usepackage[USenglish]{babel}
\usepackage{csquotes}
\AtEveryBibitem{%
  \clearlist{language}%
  \clearfield{note}%
  \clearfield{urldate}%
}
\usepackage[margin=1in]{geometry}
\usepackage{authblk}
\usepackage{fancyhdr}
\usepackage{times}
\usepackage[parfill]{parskip}

\title{\textsc{\ttitle}}
\author[1]{Ryan Steed\thanks{Email: \texttt{ryansteed@cmu.edu}.}}
\author[2]{Diana Qing}
\author[1]{Zhiwei Steven Wu}
\affil[1]{\emph{Carnegie Mellon University}}
\affil[2]{\emph{University of California, Berkeley}}
\date{\ifarxiv\else\today\fi}

\begin{document}

\onehalfspacing

\begin{titlepage}
    \maketitle
    
    \begin{abstract}
        \input{inserts/abstract.tex}
    \end{abstract}
    \setcounter{page}{0}
    \thispagestyle{empty}
\end{titlepage}

\pagestyle{fancy}
\setlength{\headheight}{14.49998pt}
\fancyhead[l]{PREPRINT}

\clearpage

\input{inserts/body}

\section*{Acknowledgments}
\input{inserts/acknowledgments}

\section*{Responsible Disclosure}
\input{inserts/responsible-disclosure}

\singlespacing
\printbibliography

\clearpage
\singlespacing
\appendix
\input{inserts/appendix.tex}

\end{document}

%% file: inserts/preamble.tex
\input{inserts/settings}

\usepackage{amsmath,amssymb,amsfonts,amsthm,graphicx,caption,xcolor,setspace,subcaption,array,booktabs,microtype}
\usepackage{hyperref}
\usepackage{mathtools}
\usepackage{enumitem}
\usepackage[ruled, lined, linesnumbered, commentsnumbered, longend]{algorithm2e}
\usepackage[USenglish]{babel}
\usepackage{csquotes}
\usepackage{fancyhdr}

\newcommand{\ind}{\mathbb{I}}

\usepackage{color-edits}
\addauthor{sw}{blue}

\makeatletter
\newcommand{\inlinesection}{\@ifstar{\@inlinesectionstar}{\@inlinesectionnostar}}
\newcommand{\@inlinesectionstar}[1]{\ifhdsr\subsubsection*{#1}\else\textbf{#1}.\fi}
\newcommand{\@inlinesectionnostar}[1]{\@inlinesectionstar{#1}\phantomsection\addcontentsline{toc}{subsubsection}{#1}}
\makeatother

\newcommand{\census}{\textsc{Census SF1}}
\newcommand{\hud}{\textsc{HUD Picture}}
\newcommand{\pums}{\textsc{PUMS}}
\newcommand{\ppmf}{\textsc{PPMF}}
\newcommand{\dpdemo}{\textsc{DP SF1}}

\DeclareMathOperator*{\argmin}{arg\,min}

\makeatletter
\newcommand*{\inlineequation}[2][]{%
  \begingroup
    \refstepcounter{equation}%
    \ifx\\#1\\%
    \else
      \label{#1}%
    \fi
    \relpenalty=10000 %
    \binoppenalty=10000 %
    \ensuremath{%
      #2%
    }%
    ~\@eqnnum
  \endgroup
}
\makeatother

\input{inserts/numbers}

%% file: inserts/settings.tex
\newcommand{\ttitle}{Quantifying Privacy Risks of Public Statistics to Residents of Subsidized Housing}

\newif\iftodos
\todostrue

\newif\ifarxiv
\arxivfalse

\newif\ifhdsr
\hdsrfalse

\todosfalse
\arxivtrue

%% file: inserts/numbers.tex
\newcommand{\numrecs}{100} %
\newcommand{\numrecsmin}{10} %
\newcommand{\timerecmatrix}{a few hours}
\newcommand{\timereclinear}{around 72 hours}

\newcommand{\numblocksbrrounding}{4,732} %
\newcommand{\hudresponserate}{97\%} %

\newcommand{\numblocks}{10,224}

\newcommand{\numsechhs}{732,757}

\newcommand{\ipblocks}{1,004} %
\newcommand{\ipblockspercent}{9.8\%}

\newcommand{\ipblocksufifteen}{62}
\newcommand{\ipblocksuthirty}{270}

\newcommand{\dpipinvalidpercent}{around 6\%}

\newcommand{\sharedipblocks}{778}
\newcommand{\dpipblocks}{2,205}

\newcommand{\synthalpha}{10^{-4}}
\newcommand{\synthbaselinesamplesize}{20\%}

\newcommand{\synthpercentsec}{47\%} %
\newcommand{\synthsechhs}{731,739}
\newcommand{\synthpercentviolationsec}{3.4\%} %
\newcommand{\synthviohhs}{24,767}
\newcommand{\synthblocksviolation}{4,590}
\newcommand{\synthblocksviolationpercent}{44.9\%}
\newcommand{\synthipblocks}{2,391}
\newcommand{\synthipblocksdetectionrate}{23\%} %
\newcommand{\synthipblocksprecision}{100\%}
\newcommand{\synthipblocksrecall}{52\%}

\newcommand{\synthprecisionhudsecknown}{66\%}
\newcommand{\synthprecisionhudsecknownBaseline}{4.6\%}
\newcommand{\synthvioratesecknown}{3.4\%} %
\newcommand{\synthrecallhudsecknown}{21\%}

\newcommand{\synthhhshudsecknownuniq}{27,853}
\newcommand{\synthviohhshudsecknownuniq}{4,677}
\newcommand{\synthprecisionhudsecknownuniq}{63\%}

\newcommand{\synthprecisionhudsecknownuniqDPDAS}{37\%}

\newcommand{\synthrecallhudsecknownuniq}{11\%}

\newcommand{\synthhhshhsecknownuniq}{17,791}
\newcommand{\synthviohhshhsecknownuniq}{4,478}
\newcommand{\synthprecisionhhsecknownuniq}{63\%}
\newcommand{\synthrecallhhsecknownuniq}{10\%}

\newcommand{\synthhhshhsecunknownuniq}{24,662}
\newcommand{\synthviohhshhsecunknownuniq}{4,478}
\newcommand{\synthprecisionhhsecunknownuniq}{40\%}

\newcommand{\synthprecisionhhsecunknownuniqDPDAS}{27\%}

\newcommand{\synthrecallhhsecunknownuniq}{10\%}

\newcommand{\synthswaprate}{10\%}

\newcommand{\synthswappedipblocks}{1,532}
\newcommand{\synthswappedipblocksdetectionrate}{15\%} %
\newcommand{\synthswappedipblocksprecision}{98\%}
\newcommand{\synthswappedipblockswrong}{28}
\newcommand{\synthswappedipblocksrecall}{33\%}

\newcommand{\numpropertiesnogeocoderresult}{3,494}
\newcommand{\numblockswithsuproperties}{1,713}
\newcommand{\numblockswithGQ}{2,476}
\newcommand{\numblocksoccupiedhhexceed}{6,433}
\newcommand{\numblockspeopletotalexceed}{444}

\newcommand{\synthdpipblocks}{6,703}
\newcommand{\synthdpipblocksdetectionrate}{66\%} %
\newcommand{\synthdpipblocksprecision}{56\%}
\newcommand{\synthdpipblocksrecall}{84\%}

\newcommand{\solvarZeroBlocks}{16}
\newcommand{\solvarZeroHouseholds}{476}

\newcommand{\solvarZeroHouseholdsVio}{none}

\newcommand{\solvarSimpleZeroBlocks}{344}
\newcommand{\solvarSimpleZeroHouseholds}{20,666}

\newcommand{\solvarSimpleZeroHouseholdsUniqSubset}{453}

\newcommand{\solvarSimpleZeroHouseholdsVioUniq}{170}
\newcommand{\solvarSimpleZeroHouseholdsSubsidized}{17,904}
\newcommand{\solvarSimpleZeroHouseholdsSubsidizedUniqSubset}{240}
\newcommand{\solvarSimpleFivePerBlocks}{921}
\newcommand{\solvarSimpleFivePerHouseholds}{104,324}

\newcommand{\solvarSimpleFivePerHouseholdsVioUniq}{804}

%% file: inserts/abstract.tex
As the U.S. Census Bureau implements its controversial new disclosure avoidance system, researchers and policymakers debate the necessity of new privacy protections for public statistics. With experiments on both public statistics and synthetic microdata, we explore a particular privacy concern: respondents in subsidized housing may deliberately not mention unauthorized children and other household members for fear of being discovered and evicted. By combining public statistics from the Decennial Census and the Department of Housing and Urban Development, we demonstrate a simple, inexpensive reconstruction attack that could identify subsidized households living in violation of occupancy guidelines in 2010. Experiments on synthetic data suggest that a random swapping mechanism similar to the Census Bureau's 2010 disclosure avoidance measures does not significantly reduce the precision of this attack, while a differentially private mechanism similar to the 2020 disclosure avoidance system does. Our results provide a valuable example for policymakers seeking trustworthy public statistics.

%% file: inserts/body.tex
\section{Introduction}
In response to concerns about reconstruction-abetted attacks on census participants' confidential responses \citep{abowd_declaration_2021, garfinkel_understanding_2018, hawes_implementing_2020}, the United States Census Bureau updated its disclosure avoidance system for the 2020 Decennial Census to adhere to differential privacy (DP), a formal privacy standard \citep{abowd_2020_2022, dinur_revealing_2003, dwork_calibrating_2006}.
Some question whether the new protections are effective or even necessary \citep{ruggles_differential_2019, kenny_use_2021}---though Title 13 requires the Census Bureau to keep individuals' responses confidential, detractors suggest that differential privacy ``goes above and beyond'' legal precedent \citep{mervis_researchers_2019}. But Census Bureau officials and scholars argue that confidentiality is more than a legal formality---respondents may be less likely to respond truthfully if they are worried about how their responses could be used against them \citep{abowd_confidentiality_2023, sullivan_coming_2020, ohare_undercount_2015}. Meanwhile, modern computing and commercial datasets empower attackers to reconstruct and re-identify individual responses from public statistics \citep{dinur_revealing_2003}---a possibility that may further deter respondents.

We investigate one specific privacy concern: that block-level Decennial Census summary statistics could be used to help reconstruct and re-identify households living in violation of their leases. In a 2020 workshop on privacy and census participation, researcher danah boyd recounted Section 8 public housing residents' fears that if they provided full information to city or census workers about who is living in their homes, they might be evicted \citep[p. 130]{boyd_panel_2020}:
\begin{quote}
    These are people who are concerned about being subject to constant surveillance: video surveillance in the buildings, concerns about surveillance of electric and water utility usage. They fear that the information that they provide might be directly used against them, but the prospect of being re-identified from public statistical information is terrifying as well.
\end{quote}
These concerns have a direct impact on data quality: \citet{ohare_undercount_2015} cites multiple reports of respondents deliberately not mentioning their children on census forms for fear of reprisal from landlords, immigration and housing agencies, social services agencies, or custodial parents.

We conduct a series of empirical experiments to explore and quantify the risk that public statistics could reveal households living in violation of occupancy limits in subsidized housing. The attacks leverage a small set of public statistics---on household size, race, ethnicity, and age---available from the 2010 Decennial Census and from Department of Housing and Urban Development administrative data. Our results suggest that an attacker could solve simple integer programs to discover as many as {\ipblocks} census blocks with households living in violation of housing rules (Table~\ref{tab:blocks}) and reconstruct the recorded characteristics of those households---often with very high confidence---within seconds per block on a standard laptop. Experiments on simulated microdata suggest these reconstructions could be used to re-identify households in violation---including households unique in their blocks---at rates much better than random chance.
We also applied a simple, random swapping procedure---similar to the one used by the Census Bureau in 2010 \citep{abowd_confidentiality_2023}---to synthetic data and found that swapping has little effect on an attacker's ability to precisely re-identify blocks and unique households in violation, while a differentially private mechanism similar to the one used for the 2020 Decennial Census reduces precision to no better than a baseline inference from public-use microdata.

The possibility that responses to the census and other government surveys may be linked back to sensitive information about households could put vulnerable groups at risk and deter them from participating fully in vital public statistics. The attack demonstrated here may not be used in practice---even if an attacker can successfully reconstruct the confidential responses, those responses, and whatever external data the attacker uses to re-identify them, may contain errors that make it more difficult to identify real households living in violation of housing rules \citep{abowd_2010_2023}. Realistically, this information may be more easily discovered through surveillance systems or other means \citep{macmillan_eyes_2023}. Still, the possibility of reconstruction---and the possibility of other, more sophisticated attacks related to housing and households---is of particular concern for census respondents and for the federal agencies tasked with keeping their information confidential.\footnote{CIPSEA in particular limits disclosure of information that could be used ``in conjunction with other data elements to reasonably infer the identity of a respondent. For example, data elements such as a combination of gender, race, date of birth, geographic indicators, or other descriptors may be used to identify an individual respondent'' \citep{office_of_management_and_budget_implementation_2007}.} Our work provides an empirical demonstration of the relative effectiveness of various privacy protection techniques and an example for policymakers and public officials designing future privacy protections. 

\subsection*{Related Work}
Differential privacy provides an appealing accounting system for privacy loss with formal guarantees about individual disclosure risk, but empirical demonstrations are also useful for communicating the immediate consequences of new privacy protections and may be more convincing to certain stakeholders \citep{nanayakkara_whats_2022}. While there are many empirical studies on the effects of census privacy protections on data quality and utility \citep{kenny_use_2021, brummet_effect_2022, asquith_assessing_2022, ruggles_differential_2019, steed_policy_2022, christ_differential_2022} and of the economic downsides to privacy more generally \citep[see][]{acquisti_economics_2023}, there are fewer empirical studies of privacy risk to census participants.

The most notable is a reconstruction attack conducted by Census Bureau scientists in which a portion of published tables from the 2010 Decennial Census was sufficient to reconstruct and (when linked with commercial data) re-identify the personal characteristics of millions of individuals \citep{abowd_2010_2023}. Our focus on a specific harm to respondents was inspired by another census linkage attack that could possibly be used to re-identify transgender children in Texas \citep{flaxman_risk_2025}. Unlike these two examples, our work also leverages statistics published by a second government agency---the Department of Housing and Urban Development---that, as far as we can tell, protects respondent data only with property-level aggregation and suppression of small properties. Our work also draws on a long line of empirical and theoretical research on reconstruction, re-identification, and database inference attacks \citep{dinur_revealing_2003, cohen_linear_2020, narayanan_how_2007, sweeney_simple_2000, dick_confidence-ranked_2023, dwork_price_2007}.

\section{Study Design}
Tenants in subsidized housing are assigned to units on the assumption that their households meet occupancy standards, which may even be included explicitly in their lease \citep[Chs. 3 \& 5]{us_department_of_housing_and_urban_development_hud_2013}.
Though exceptions are possible, HUD suggests that at a minimum property managers adopt a ``two heartbeats per room'' policy---no more than two persons should be required to occupy a bedroom \citep{us_department_of_housing_and_urban_development_tenant_1987, us_department_of_housing_and_urban_development_hud_2013}.
Respondents may rightly fear that if a landlord, property manager, or housing official discovers that their household violates this occupancy standard, they will be evicted from their homes---indeed, sophisticated surveillance systems installed by some housing agencies have already been used to provide evidence to evict public housing residents \citep{macmillan_eyes_2023}.

\inlinesection*{Public data.} We evaluate privacy risk to {\numsechhs} subsidized households in {\numblocks} census blocks in the 50 states, the District of Columbia, and Puerto Rico.
We link two publicly available datasets: (1) block-level statistics from the 2010 Summary File 1 ({\census}) summarizing data collected in the 2010 Decennial Census \citep{us_census_bureau_2010_2011}; and (2) property-level administrative statistics \citep{us_department_of_housing_and_urban_development_picture_2010} from the U.S. Department of Housing \& Urban Development ({\hud}) on subsidized housing units leased to low-income households in 2010 under Section 8, Section 236, and other housing programs. (2020 versions of these statistics are also publicly available.) The attack leverages just a small set of Person and Household tables from the 2010 Summary File 1, including the number of households, the number of housing units, the number of households that identified as each race and ethnic group, the number of children (those younger than 18), and the number of households of sizes 1 through 7+, all at the block level. From the {\hud} statistics, we use the number of housing units, the number of households, the percentage of housing units with 0 or 1 bedroom, 2 bedrooms, or 3+ bedrooms,\footnote{Bedroom counts are reported as a percentage of total units in each property; we use these proportions to compute the number units with each bedroom count, rounding down and allowing the remaining units to contain any number of bedrooms.} and information on what percentage of households include a certain race, ethnicity, or age group.

We link the two datasets by matching property addresses with census blocks and aggregating property statistics at the block level (see Appendix~\ref{app:link}). We excluded {\numblockswithsuproperties} properties without valid addresses and {\numblocksoccupiedhhexceed} blocks where {\hud} reports more households or persons than {\census} (Appendix~\ref{app:link}); this may occur due to errors in the property addresses or when properties span multiple blocks. With additional data on residential buildings, an attacker may be able to better match households to census blocks; here, we simply exclude properties which obviously overlap. HUD suppresses properties with less than 11 subsidized households \citep{us_department_of_housing_and_urban_development_picture_2010}, so all the blocks we examine have at least 11 subsidized households; to simplify our approach, we also ignore blocks with group quarters. Later, using the same procedure, we also experiment with statistics from the differentially private 2010 Demographic and Housing Characteristics (DHC) Demonstration Data ({\dpdemo}) released in August 2022 \citep{us_census_bureau_development_2022}.

\inlinesection*{Approach.} We imagine an attacker interested in learning which subsidized households are living in violation of occupancy standards. To find violations of the ``two heartbeats per room'' policy, the attacker aims to match household sizes (based on statistics from {\census}) to bedroom counts (based on statistics from {\hud}). For a multiset of households $D$ (represented by rows of discrete household attributes) from the confidential microdata for a given census block, the attacker seeks to: \textbf{1) reconstruct (\S\ref{sec:attacks-recon})} a multiset of microdata records $D'$ using the two sets of statistics $Q(D) = (Q_{\census}(D_{\census}), Q_{\hud}(D_{\hud}))$
and identify blocks in which at least one violation of housing guidelines must exist; \textbf{2) re-identify (\S\ref{sec:attacks-reid})} reconstructed households in violation of housing guidelines in those blocks.

\inlinesection*{Synthetic validation data.} Because the microdata used to produce these statistics are confidential, we also validate the attacks on statistics produced from synthetic data (see \S\ref{sec:validation})---simulated using the privacy-protected microdata files (\ppmf) from the April 2023 vintage of the DHC demonstration data \citep{us_census_bureau_just_2023}.

\section{Attack on 2010 public statistics}
\label{sec:results-sf1}
\subsection{Reconstruction attack}
\label{sec:attacks-recon}
\inlinesection*{Detecting blocks with households in violation.} We specify an integer program with constraints on the subsidized status, size, householder race/ethnicity, and age composition of the subsidized households in a block using summary statistics from {\census} and {\hud}. For each block, an attacker could search for solutions to $Q(D') \approx Q(D)$,\footnote{We relax this equality for some queries to allow for missing information. Also, the attacker assumes that these statistics are produced from the same population---a set of joined microdata $D$---but in reality the statistics are produced from separate sets of microdata $D_{\census}$ and $D_{\hud}$. We relax constraints to allow for discrepancies between the two sets of statistics, but the attack is still subject to response error from both sets of data. For a full description of the exact statistics and constraints used, see Appendix~\ref{app:ip}.} where the variables to be reconstructed in $D'$ are the demographic characteristics, household size, and bedroom count of each household in the block. To simply identify blocks with at least one household in violation, an attacker could additionally constrain the size of each subsidized household by the number of bedrooms in the unit according to HUD's ``two heartbeats per room'' occupancy guideline. In this case, if $V' \subseteq D'$ are the households with attributes in violation of this rule, the attacker attempts to solve
\begin{align}
    &Q(D') \approx Q(D) \label{eq:csp}\\
    &|V'| = 0.\nonumber
\end{align}

We solve this constraint satisfaction program with Gurobi \citep{gurobi_optimization_llc_gurobi_2023}, identifying blocks for which no possible reconstruction of the subsidized households could satisfy these occupancy limits.\footnote{In Gurobi, we write the occupancy limits constraint in two parts: one constraint requires that the number of 0 or 1 bedroom units with more than 2 occupants is zero, and another requires that the number of 2 bedroom units with more than 4 occupants is zero.} It takes only {\timerecmatrix} to solve this simple constraint satisfaction program for all {\numblocks} blocks on a server with 32 8-core, 2-thread, 2.9GHz CPUs and 256GB RAM.

\begin{figure}[!t]
    \centering
    \includegraphics[width=\linewidth]{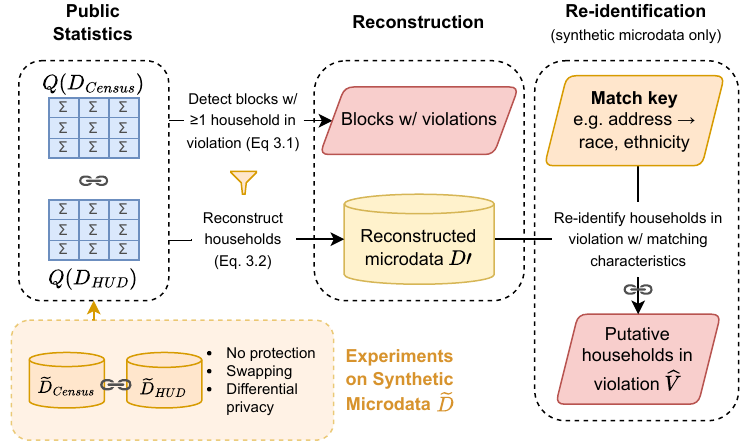}
    \caption{Overview of experimental approach. A simulated re-identification attack is conducted on statistics computed from synthetic microdata, using a subset of the synthetic microdata as a match key (a subset of household characteristics mapped to an identifier such as an address). In practice, the attacker must obtain this match key from a public, e.g. commercial, source. The attacker may reduce computation and increase precision of re-identification by only targeting blocks with at least one violation (Eq.~\ref{eq:csp}).}
    \label{fig:diagram}
\end{figure}

\inlinesection*{Reconstructing households in violation.} For the vast majority of blocks, there are multiple reconstructions $D'$ that satisfy $Q(D') \approx Q(D)$. To answer more detailed questions about which households are likely to be in violation of housing guidelines, an attacker may use prior information to select more likely reconstructions to later re-identify. As an example, we construct an empirical likelihood function from the 2010 Public Use Microdata Samples ({\pums}) \citep{us_census_bureau_2010_2014}, a 10\% sample of households in each U.S. state, and to reduce computation consider only household configurations that appear at least once in {\pums}. For each block in a given U.S. state, we solve the integer program
\begin{align}
    \argmin_{D'}\quad &- \sum_{d' \in D'} \log \hat{p}_{\text{census}}(d') \label{eq:emle}\\
    \text{subject to}\quad &Q(D') \approx Q(D) \nonumber\\
    &\hat{p}_{\text{census}}(d') > 0 \quad \forall d' \in D', \nonumber
\end{align}
where $\hat{p}_{\text{census}}(d')$ is the proportion of households in the state with attributes matching household $d' \in D'$. Note that {\pums} only includes {\census} variables, and so does not include information about bedroom counts or subsidized status. Solving for the optimal solution to Eq.~\ref{eq:emle} with Gurobi takes only slightly longer than solving the simple constraint satisfaction problem. However, solving for the top $t$ optimal solutions with Gurobi takes much longer---{\timereclinear} for all {\numblocks} blocks when $t=\numrecs$.

\inlinesection*{Solution variability.} Without access to the underlying microdata, an attacker cannot definitively confirm their reconstructions of households in violation. However, an attacker can compute the \emph{solution variability} of a given reconstruction, an upper bound on the number of reconstructed households that may be incorrect \citep{abowd_2010_2023}. Adapting from \citet{abowd_2010_2023}, we define the solution variability of a feasible reconstruction of households $D^{*}$ or households in violation $V^{*} \subseteq D^{*}$ in a given block with $L_1$ (Manhattan) distance:
\begin{align}
    solvar(D^{*}) = &\max_{\{D' | Q(D') \approx Q(D)\}} L_1\{\text{Hist}_M(D^{*}), \text{Hist}_M(D')\} \label{eq:solvar}
\end{align}
where $\text{Hist}_M(D')$ is the fully saturated (including all possible configurations of a set of household attributes $M$) histogram  in reconstruction $D'$. When $solvar$ is zero, the attacker knows that $D^{*}$ is the \emph{only} feasible reconstruction of the household attributes in $M$. Since the number of households in a reconstruction $D'$ is fixed, $solvar(D^{*})/2|D^{*}|$ is an upper bound on the \emph{percentage} of records in a feasible reconstruction $D^{*}$ that could differ in attributes $M$ in any other possible reconstruction.
\citet{abowd_2010_2023} additionally show with the triangle inequality that any attacker's $solvar(D^{*})$ is within a factor of 2 of any other feasible $solvar$.

By default, $M$ contains all the household attributes in the {\census} and {\hud} statistics. However, an attacker may not need to know the granular race, ethnicity, and age of a household's members to re-identify it; later, we also report an attacker's confidence in fewer and less precise details about a household by computing $solvar$ using the marginal histogram for less granular (binned) attributes.

\subsection{Results from 2010 public data}
\inlinesection*{Attack on 2010 Summary File 1.} After running the reconstruction attack (\S\ref{sec:attacks-recon}) on all {\numblocks} blocks with subsidized housing, we find that by solving a simple constraint satisfaction problem (Eq.~\ref{eq:csp}) with queries from {\census} and {\hud}, an attacker could determine that {\ipblocks} blocks ({\ipblockspercent} of blocks with subsidized properties) contain at least one household in violation of the ``two heartbeats'' occupancy limit (not accounting for swapping and data error; see \S\ref{sec:limitations}).
With just this information, it may already be easy to re-identify households are in violation---{\ipblocksufifteen} of these blocks have less than 15 households in a subsidized property; {\ipblocksuthirty} have less than 30 subsidized households (Figure~\ref{fig:violations-s8reported}).

Most of these reconstructions are not unique solutions---in fact, for the majority of blocks, there is another feasible reconstruction where all the records differ ($solvar(D^{*})/2|D^{*}| = 1$), likely due to the small number of statistics used in this attack.
Only {\solvarZeroBlocks} reconstructed blocks (containing {\solvarZeroHouseholds} reconstructed households, {\solvarZeroHouseholdsVio} of which violate occupancy limits) have \emph{only one} feasible reconstruction (zero solution variability) over all the attributes we reconstructed. These are reconstructed blocks the attacker is certain match the 2010 confidential microdata post-swapping and other disclosure avoidance measures. (In \S\ref{sec:validation}, we use synthetic data to also simulate the precision of this attack relative to the confidential microdata \emph{before} disclosure avoidance.)

\begin{figure}
    \centering
    \begin{subfigure}[t]{0.49\linewidth}
        \centering
        \includegraphics[width=\linewidth]{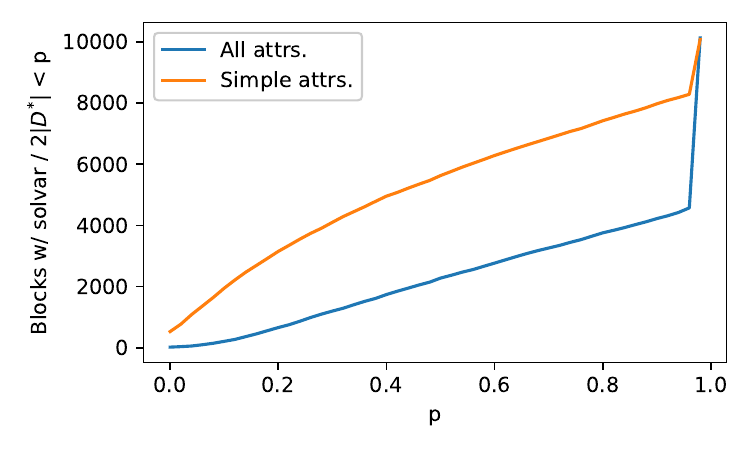}
        \caption{Distribution over blocks.}
    \end{subfigure}
    \begin{subfigure}[t]{0.49\linewidth}
        \centering
        \includegraphics[width=\linewidth]{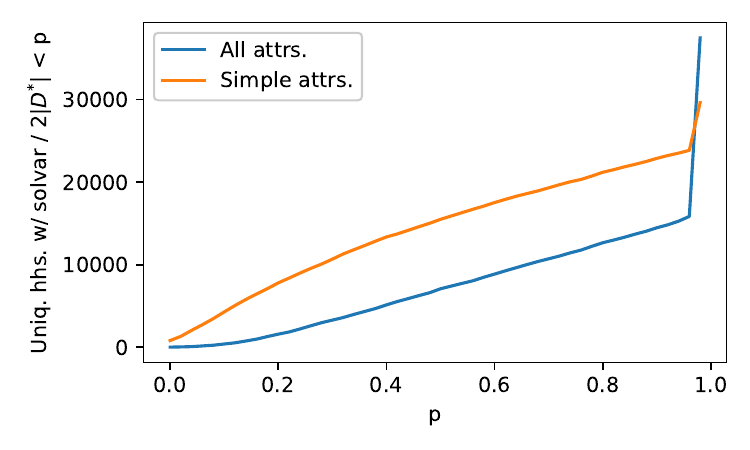}
        \caption{Distribution over unique households.}
    \end{subfigure}
    \caption{Cumulative distribution of solution variability as a percentage of reconstructed records for each block reconstructed from public statistics in {\census} and {\hud}. By default, we compute the $solvar$ of all reconstructed households over all possible attributes. We also compute $solvar$ and household uniqueness with a \emph{simpler} set of attributes: household size, bedroom count, presence of a white, non-Hispanic householder, and presence of a child.}
    \label{fig:solvar}
\end{figure}

However, solution variability is much lower if the attacker does not need granular age, race, and ethnicity attributes for re-identification. There are {\solvarSimpleZeroBlocks} reconstructed blocks with \emph{only one} feasible reconstruction (zero variability) of just household size, bedrooms, presence of a white, non-Hispanic householder, and presence of a child. These blocks contain {\solvarSimpleZeroHouseholds} perfectly reconstructed households (including {\solvarSimpleZeroHouseholdsSubsidized} subsidized households). {\solvarSimpleZeroHouseholdsUniqSubset} of these households (including {\solvarSimpleZeroHouseholdsSubsidizedUniqSubset} subsidized households) are unique in their block in these variables. {\solvarSimpleZeroHouseholdsVioUniq} are both unique and in violation of occupancy limits. And there are even more blocks with \emph{only a few} feasible reconstructions---an additional {\solvarSimpleFivePerBlocks} reconstructed blocks (with {\solvarSimpleFivePerHouseholds} reconstructed households, {\solvarSimpleFivePerHouseholdsVioUniq} of which are unique and in violation), for example, may differ in no more than 5\% of records in this simplified set of attributes ($solvar(D^{*})/2|D^{*}| < 0.05$). Figure~\ref{fig:solvar} shows the full cumulative distribution of solution variability over blocks and households.

\inlinesection*{Attack on DP Demonstration Data.} To comparatively test the effects of the full production disclosure avoidance system, we also ran attacks on the the 2010 Demographic and Housing Characteristics (DHC) Demonstration Data ({\dpdemo}) released in August 2022 \citep{us_census_bureau_development_2022}.
This data can be viewed as the statistics that would have resulted had the Census Bureau used the 2020 disclosure avoidance system \citep{abowd_2020_2022} to produce the 2010 {\census} instead of swapping and other traditional disclosure avoidance methods \citep{mckenna_disclosure_2018}.
With the noisy {\dpdemo} statistics, an attacker would no longer be able to reconstruct valid microdata for {\dpipinvalidpercent} of blocks with the same attack (Eq.~\ref{eq:csp}) and in the remainder, the number of blocks identified as having violations inflates drastically from {\sharedipblocks} to {\dpipblocks}. To include all blocks, the attacker could instead reconstruct by minimizing query error (Appendix~\ref{app:soft}), but in our setting still fails to identify most of the original blocks with violations.

\section{Validation with synthetic statistics}
\label{sec:validation}
Because the ``ground truth'' microdata used to produce the 2010 {\census} are confidential, we evaluate the possibility of re-identification using a set of simulated ground truth microdata adapted from the 2010 demonstration privacy-preserving microdata files (PPMF) (Appendix~\ref{app:simulation}). Of course, the confidential microdata used to produce the original {\census} were not completely unprotected---certain tables were produced with swapping and other opaque disclosure avoidance measures \citep{abowd_confidentiality_2023, mckenna_disclosure_2018}. To compare different levels of protection, we ran the same reconstruction attack (\S\ref{sec:attacks-recon}) on statistics $Q(\tilde{D})$ produced from synthetic microdata $\tilde{D}$ under several different conditions (\S\ref{sec:results-synth}): 1) no privacy protections; 2) random swapping, approximating the original 2010 Decennial Census release; and 3) differential privacy, approximating the disclosure avoidance system developed to protect the 2020 Decennial Census \citep{abowd_2020_2022} with the discrete Gaussian mechanism \citep{canonne_discrete_2022}. Then, we re-identified synthetic households in violation using a hypothetical set of (perfect) commercial data drawn from the synthetic microdata (\S\ref{sec:attacks-reid}).

\subsection{Re-identification attack}
\label{sec:attacks-reid}
\inlinesection*{Re-identifying reconstructed households with a linkage attack.} To measure the potential effectiveness of these attacks for re-identifying the households in violation $V$ in a given block reconstruction $D$, we imagine an attacker attempting to re-identify reconstructed households in violation by linking them with a set of identified data (e.g., a list of householders with their race, ethnicity, and address obtained commercially). Suppose the attacker has access to such a \emph{match key} $D_\mathcal{S} \subseteq \mathcal{S}$---a set of partial microdata with identifiers including only a subset $\mathcal{S}$ of the household attributes (e.g., from commercial data).
Similar to \citet{dick_confidence-ranked_2022}, the attacker constructs a ranking $R_{\mathcal{S}}'$ ordering the possible household configurations by how often a household in violation in the $t>\numrecsmin$\footnote{For the experiments that follow, $t=\numrecs$ by default. For differential privacy, some of the lower values of $\rho$ we tested result in a high number of reconstructed blocks in putative violation, so for ease of computation we dynamically lower the number of reconstructions used to compute performance metrics to $t>\numrecsmin$ for just those treatments.} most likely reconstructions of a block has configuration $R_i' \in \mathcal{S}$. The attacker then finds \emph{putative violations} $\hat{V} = D_\mathcal{S} \cap R_{\mathcal{S}\;{1:k}}'$ by joining their match key with the top $k$ household configurations in $R_{\mathcal{S}}'$.

In our validation experiments with synthetic data (\S\ref{sec:validation}), we measure the effectiveness of this method for identifying households actually in violation with precision and recall:
\begin{align}
    \textsc{Precision}_{V,k}(R_{\mathcal{S}}') &\coloneqq \frac{|V \cap \hat{V}|}{|\hat{V}|}\\
    \textsc{Recall}_{V,k}(R_{\mathcal{S}}') &\coloneqq \frac{|V \cap \hat{V}|}{|V|}.
\end{align}
Precision is the fraction of putative violations that can be matched to a true violation; recall is the fraction of true violations successfully identified. We also evaluate household precision and recall over population uniques---households that are unique within their blocks in the attackers' identified partial microdata (the match key) \citep{elamir_record_2006}. To increase precision, we imagine the attacker first detects blocks with at least one household in violation (Eq.~\ref{eq:csp}), then searches for households in violation only in those blocks.

\inlinesection*{Sampling attack (strong distributional baseline).} \citet{dick_confidence-ranked_2023} note that if the dataset distribution is very low entropy, an attacker could match reconstructed households to real households at a high rate by simply guessing rows that are likely under the data distribution. As a baseline for comparison, we imagine an attacker with previous knowledge of the configurations of households most likely to be in violation. Suppose the attacker has access to the empirical distribution of a {\synthbaselinesamplesize} microdata sample from each geographic state in the ground truth synthetic data $\tilde{D}$ (similar to {\pums}). Then the attacker constructs the ranking $R_\mathcal{S}$ by how frequently a household with each attribute combination $R_i \in \mathcal{S}$ appears in violation in the empirical sample. This imaginary attack serves as an unrealistically strong baseline for empirical inference from a microdata sample. In practice, such detailed information about the distribution of violating households is likely not available. In this case, the attack is especially unrealistic because {\pums} includes neither subsidized status nor bedroom counts, the two variables necessary to identify violations of occupancy limits.

\subsection{Results from synthetic data}
\label{sec:results-synth}
\inlinesection*{Attacks on statistics with no privacy protections.} Our experiments on synthetic data suggest an attacker can use reconstructions to identify blocks and households in violation at rates much higher than chance. In our synthetic microdata, {\synthsechhs} of households live in subsidized properties ({\synthpercentsec} of all households, same as in the original {\hud} and {\census}). {\synthviohhs} ({\synthpercentviolationsec}) of these synthetic subsidized households are in violation of the occupancy limits. Violations are spread fairly evenly across blocks in our simulation---{\synthblocksviolationpercent} of blocks have at least one household in violation. Running the reconstruction attack on statistics produced from our synthetic microdata, an attacker could identify {\synthipblocksrecall} (block-level recall) of these blocks with ground truth violations, much better than random chance (Table~\ref{tab:blocks}). We also found a strong correlation between the number of households in violation and the number of reconstructed households in violation (putative violations) in each block (Appendix~Figure~\ref{fig:violation-correlation}).

\begin{table}[!t]
        \centering
        \begin{tabular}{rccc}
            \hline
             & No protections & Swapping (10\%) 
             & DP (DAS budget) \\\hline
             Blocks with $\geq 1$ violation & {\synthblocksviolation} ({\synthblocksviolationpercent}) &
             {\synthblocksviolation} ({\synthblocksviolationpercent}) &
             {\synthblocksviolation} ({\synthblocksviolationpercent})  \\
             Putative blocks with $\geq 1$ violation & {\synthipblocks} ({\synthipblocksdetectionrate}) &
             {\synthswappedipblocks} ({\synthswappedipblocksdetectionrate}) &
             {\synthdpipblocks} ({\synthdpipblocksdetectionrate}) \\
             Block-level precision & {\synthipblocksprecision} & {\synthswappedipblocksprecision} &
             {\synthdpipblocksprecision}  \\
             Block-level recall & {\synthipblocksrecall} & {\synthswappedipblocksrecall} &
             {\synthdpipblocksrecall} \\
             \hline
        \end{tabular}
        \caption{Block-level precision and recall when the attacker attempts to identify blocks with at least one household in violation (Eq.~\ref{eq:csp}) using statistics produced from {\numblocks} synthetic blocks, with and without simulated random swapping (swap rate {\synthswaprate}) or differential privacy (using an approximation of the April Census TopDown budget).
        The true violation rate is the expected precision if putative blocks in violation are selected randomly (1st row); the rate of putative (predicted) violations is the expected recall (2nd row).
        Swapping reduces the number of putative violations, but does not significantly reduce precision.
        }
    \label{tab:blocks}
\end{table}

\begin{figure}[!t]
    \centering
    \begin{subfigure}[t]{0.49\linewidth}
        \includegraphics[width=\linewidth]{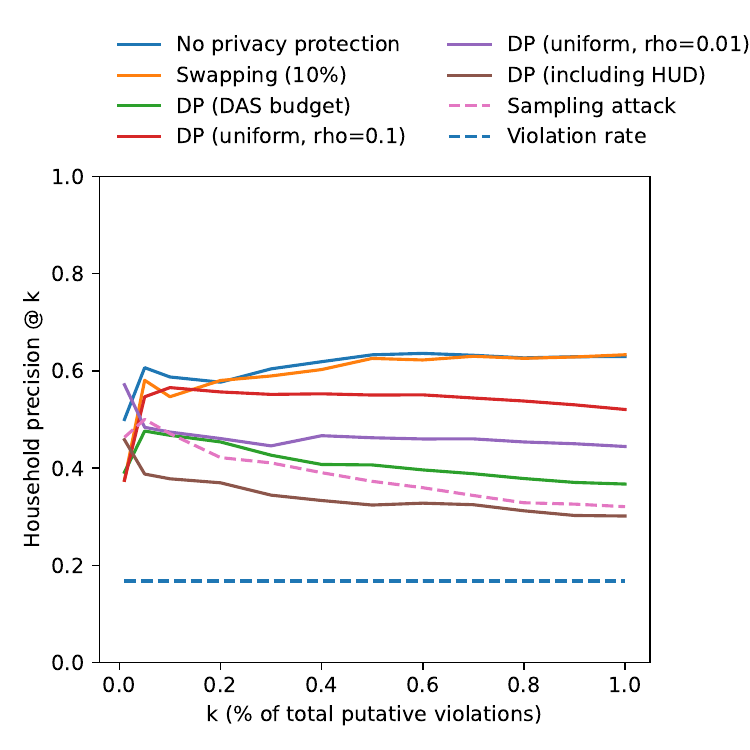}\\
        \caption{Matching with subsidized status, bedroom counts, householder race/ethnicity, and presence of a child (the {\hud} microdata); {\synthhhshudsecknownuniq} unique households.}
        \label{fig:precision-unique-hud-vars}
    \end{subfigure}
    \hfill
    \begin{subfigure}[t]{0.49\linewidth}
        \includegraphics[width=\linewidth]{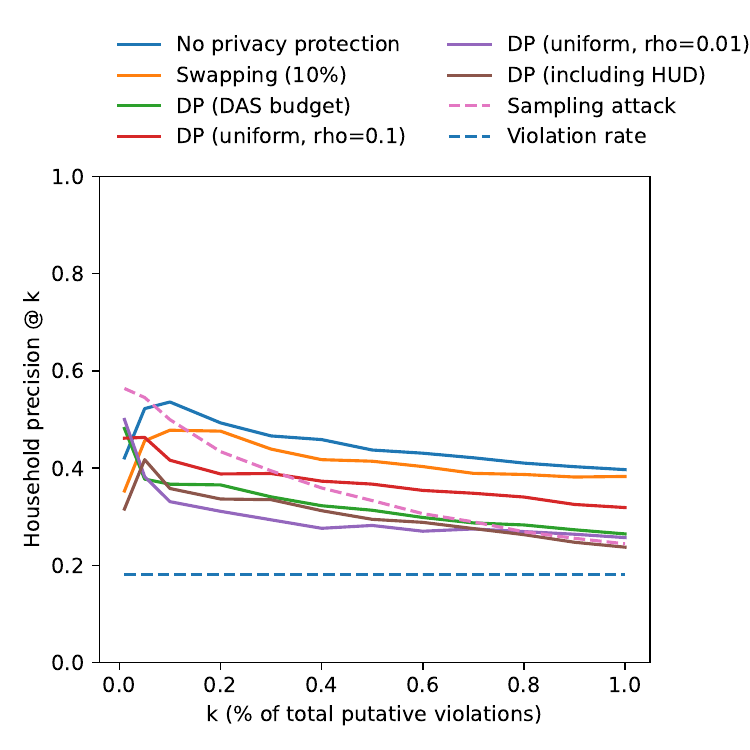}\\
        \caption{Matching with householder race/ethnicity, and presence of a child (no subsidized status or bedroom counts); {\synthhhshhsecunknownuniq} unique households.}
        \label{fig:precision-unique-hh-vars-s8unknown}
    \end{subfigure}
    \caption{$\textsc{Precision}_{V,k}(R_\mathcal{S})$ of reconstruction \& re-identification when an attacker re-identifies households in violation \emph{unique in their blocks} by matching the top $k$ most frequent putative violations (reconstructed households in violation) with a hypothetical identified set of real households with variables $R_\mathcal{S}$ (the match key). The violation rate is the expected precision if putative violations are randomly selected.}
    \label{fig:precision-unique}
\end{figure}

Moreover, the attributes of putative households in violation (in just the blocks known to have violations) could be used to re-identify real households in violation. Matching reconstructed households to identified records of bedroom count, householder race/ethnicity, and presence of a child---possible for an attacker with access to the {\hud} records, for example---could identify {\synthrecallhudsecknown} of households in violation (recall) at a rate of {\synthprecisionhudsecknown} precision or higher (in predictions on the subsidized households, where the simulated violation rate is {\synthvioratesecknown}). This exceeds the {\synthprecisionhudsecknownBaseline} precision of inference in a scenario where the attacker already knows the most frequent characteristics of violating households in the confidential microdata (Figure~\ref{fig:precision-recall-hudsecknown}, sampling attack at $k=|\hat{V}|$, when all putative violations are included). If the attacker targets only the {\synthhhshudsecknownuniq} households unique in the match key within their block, an attacker could identify {\synthrecallhudsecknownuniq} of the {\synthviohhshudsecknownuniq} unique households in violation with precision {\synthprecisionhudsecknownuniq} or higher (Figure~\ref{fig:precision-unique-hud-vars}). Similarly, matching with just the {\synthhhshhsecknownuniq} unique identifiable records of householder race/ethnicity and the presence of a child in the household---data points advertised by the data broker Experian \citep{experian_marketing_services_consumerview_2018}---an attacker could identify {\synthrecallhhsecknownuniq} of the {\synthviohhshhsecknownuniq} unique households in violation with {\synthprecisionhhsecknownuniq} precision if subsidized status is known (Figure~\ref{fig:precision-recall-hh-s8known}). 

Precision may be more limited if the attacker does not already know which households are in subsidized properties. Not knowing which households live in a subsidized property, an attacker could identify {\synthrecallhhsecunknownuniq} of the {\synthviohhshhsecunknownuniq} unique households in violation by matching on race/ethnicity and presence of a child, with precision only {\synthprecisionhhsecunknownuniq} (Figure~\ref{fig:precision-unique-hh-vars-s8unknown}). (This information may be easy to discover, however, if the attacker is using addresses to identify households---{\hud} includes the addresses of subsidized properties.)

\inlinesection*{Attacks on statistics after swapping.} Before releasing statistics from the 2010 Decennial Census, the Census Bureau identified households considered vulnerable to re-identification, including unique households and households in less populous blocks \citep{mckenna_disclosure_2018}. Then, the agency randomly swapped a portion of those records (the swap rate) with those of nearby households with the same household size and number of adults (or, conversely, number of children) \citep{abowd_confidentiality_2023}. Since the exact targeting procedure and swap rate are kept confidential \citep{crimi_top-coding_2014}, we adopt a simplified version of this procedure following \citet{kim_effect_2015} and \citet{christ_differential_2022}. For a given state or territory, we assigned households with more unique characteristics in their tract as well as households in less populous blocks a higher chance of being swapped.\footnote{Within each state, the 0.5\% most unique households living in the least populous blocks had a 100\% chance of being swapped, the next 19.5\% of households had a 60\% chance of being swapped, the next 30\% of households had a 30\% chance of being swapped, and the last 50\% households with the least unique characteristics living in the most populous blocks had a 10\% chance of being swapped.} For each household chosen for swapping, we assigned a random swap partner from the five geographically-closest households with the same household size and number of adults (the swap key used in the 2010 Census).

Swapping at {\synthswaprate}---a moderate value within the ranges tested in prior work \citep{kim_effect_2015, christ_differential_2022}---reduced the total number of households in violation re-identified (recall) but did not reduce precision (Table~\ref{tab:blocks}). The number of blocks with putative violations dropped from {\synthipblocks} ({\synthipblocksrecall} block-level recall) to {\synthswappedipblocks} ({\synthswappedipblocksrecall} block-level recall, around twice as high as chance). But only {\synthswappedipblockswrong} of those putative blocks in violation were false flags---block-level precision was still {\synthswappedipblocksprecision}. As a result of the decreased block-level recall, household recall also decreased somewhat, but household precision changed little (Figures~\ref{fig:precision-unique},~\ref{fig:recall-unique}--\ref{fig:precision-recall-hh-s8known}).

\inlinesection*{Attacks on statistics released with differential privacy.} To estimate the impacts of differential privacy, we simulated the discrete Gaussian mechanism used in the 2020 disclosure avoidance system, generating noisy statistics
\begin{equation}
Q(D) + \mathcal{N}_{\mathbb{Z}}\left(0, \frac{1}{\mathbf{c} \rho}\right),
\end{equation}
where $c$ is the proportion of the privacy loss budget $\rho$ allocated to each query \citep{abowd_2020_2022, canonne_discrete_2022}. Our mechanism simplifies the 2020 disclosure avoidance system in two ways: first, we apply noise to only the block-level statistics we use, by reverse engineering the budget allocated to each DAS strategy query; second, we approximate the DAS post-processing with simpler bottom-up optimization (see Appendix~\ref{app:budget} for details). By default, we used the same privacy loss budget as the April 2023 DHC release: $\rho=4.96$ for person tables and $\rho=7.70$ for household tables, allocated preferentially across statistics \citep{us_census_bureau_privacy-loss_2023}. We also tried a simpler budget allocation where the global budget $\rho$ is divided \emph{uniformly} among only the queries we use. (This results in a per-query privacy loss budget orders of magnitude higher than the DAS allocations, as we only leverage a small subset of the total queries published by the Census Bureau.)

With the DAS 2020 privacy loss budget, use of this mechanism drastically reduced precision in identifying blocks with violations (Table~\ref{tab:blocks}). The number of blocks with putative violations increased drastically to {\synthdpipblocks} with {\synthdpipblocksrecall} block-level recall, somewhat higher than chance ({\synthdpipblocksdetectionrate} recall if the same number of blocks in putative violation were selected randomly). But these guesses totaled many more than the true number of blocks with violations and were very often incorrect: block-level precision dropped to {\synthdpipblocksprecision}, only slightly better than chance.
By comparison, the attack against swapping (swap rate {\synthswaprate}) was more conservative, resulting in fewer putative violations (only {\synthswappedipblocksrecall} block-level recall), but also far more precise ({\synthswappedipblocksprecision} block-level precision).

Likewise, all the uniformly distributed DP budgets we tried ($\rho \leq 0.1$) generally resulted in worse precision in identifying unique households than swapping, recall held equal (Figure~\ref{fig:precision-unique}, Figure~\ref{fig:recall-unique}).
For example, for an attacker with access to the HUD microdata, precision on unique households at $k=|\hat{V}|$ (when all putative violations are included) decreased from {\synthprecisionhudsecknownuniq} to {\synthprecisionhudsecknownuniqDPDAS}; for an attacker with just race/ethnicity and presence of a child, precision on unique households at $k=|\hat{V}|$ decreased from {\synthprecisionhhsecunknownuniq} to {\synthprecisionhhsecunknownuniqDPDAS}, only slightly better than inference from public microdata.

These experiments assume that the {\hud} statistics are not released with differential privacy---as far as we can discern from public documentation, the {\hud} statistics are currently protected only by property-level aggregation and suppression of properties with less than 11 households \citep{us_department_of_housing_and_urban_development_picture_2010}, and we are not aware of any plans to modernize these disclosure avoidance measures. However, when we tried releasing the {\hud} statistics with a differential privacy mechanism (using the same privacy loss budget as the 2020 DAS, distributed evenly across queries), household-level precision dropped even lower (close to random for some match keys; Figure~\ref{fig:precision-recall}).

\section{Limitations}
\label{sec:limitations}
How closely our results represent the real-world effectiveness of this attack depends on several assumptions. First, the attacks on the original 2010 data aim to reconstruct confidential census records that already contain numerous forms of error, including non-response \citep{khubba_national_2022}, mis-reporting \citep{bound_measurement_2001}, and collection errors \citep{us_census_bureau_measures_2020}.
Because of these data errors, even a successful reconstruction of the confidential microdata from 2010 may not always correspond to real households in violation---for example, the Census Bureau estimates that in 2010, 3.7\% of enumerations of renters were erroneous duplicates \citep{mule_census_2012}. Conversely, the Bureau estimates an omission rate of 8.5\% for renters, which could disguise households in violation from this reconstruction attack.

Second, the real-world effectiveness of this attack is likely reduced further by errors introduced when joining the {\census} and {\hud} data (e.g., due to erroneous property addresses or discrepancies in responses between the two datasets; Appendix~\ref{app:link}), which could increase false positives. Without more detailed information on the addresses of individual households, it is difficult to quantify these errors. They may be more likely in blocks with an especially high proportion of subsidized households; however, the attack can still reconstruct households with high confidence in blocks with a small percentage of subsidized households (Figure~\ref{fig:solvar-s8ratio}).

In practice, then, this attack would likely achieve lower reconstruction quality and less precision than reported here relative to the true characteristics of 2010 households. The use of additional census tables or more accurate and detailed external data---available for purchase commercially, for example---could mitigate some of these challenges and increase the risk of re-identification; less accurate commercial data could decrease it \citep{abowd_2010_2023}. Moreover, there are many public data sources that could assist re-identification: public U.S. state voter registration data, for example, commonly include name, address, exact age, gender, and race/ethnicity \citep{pults_americas_2020}.
Future work should explore further the practical efficacy of these methods.

Finally, because the microdata used to produce the 2010 {\census} are confidential, our evaluations of the effect of swapping and differential privacy rely on a synthetic baseline for comparison. The synthetic microdata we use, especially our method for simulating occupancy limit violations, may not generalize to the empirical distribution of subsidized households in the confidential census microdata, though they are based on public microdata. Also, our replications of the Census Bureau's disclosure avoidance measures are not exact approximations of the processing applied to the 2010 {\census} and DP Demonstration data, though we think they are representative of the protections applied.

What do these results suggest about other kinds of attacks against other kinds of public statistics? The datasets and constraints used in our attack are to some extent specific to subsidized housing, but our techniques and findings mirror other reconstruction and linkage studies comparing privacy mechanisms \citep{dick_confidence-ranked_2022, christ_differential_2022, abowd_2010_2023, flaxman_risk_2025}. Census Bureau scientists and other scholars have also found that similar reconstruction \citep{abowd_2010_2023} and linkage attacks \citep{christ_differential_2022, flaxman_risk_2022} were less successful against similar DP mechanisms than against swapping-based protections. However, real-world threat models can diverge widely from our setting depending on the dataset generation process and the attackers' goals and level of access \citep{cummings_attaxonomy_2024}. Future work could test other scenarios---for example, a stronger attack using more population-level auxiliary knowledge or everyday attack by a ``curious individual'' without access to auxiliary data---and examine reconstruction robustness in housing statistics more generally \citep{cummings_attaxonomy_2024}. Methods from the growing field of privacy auditing could also prove useful for more rigorously attacking DP implementations \citep{tramer_debugging_2022}.

\section{Conclusion}

This study is a motivating example for policymakers aiming to design a more trustworthy census. We center specific concerns deterring residents of public housing from participating in the census and other government surveys and show how public statistics from the Decennial Census and HUD could be combined to reveal households living in violation of HUD occupancy limits. In seconds on a standard laptop, an attacker can link 2010 Decennial Census and HUD data to detect if a given census block contains a household in violation of the standard HUD occupancy limits and generate a set of reconstructions that could be used to identify the recorded characteristics of those households. The attack leverages only a very small subset of the published Decennial Census Statistics---using more Decennial Census tables, additional external data, and more sophisticated reconstruction techniques could improve this attack even further \citep{abowd_2010_2023}.

More expansive evaluation of the 2020 disclosure avoidance system and alternative protections is needed, but our experiments on synthetic data in this simplified setting provide some evidence that the 2020 disclosure avoidance system---and similar systems deployed at other government agencies---could provide stronger protection against privacy concerns that impede trustworthy public data.

%% file: inserts/acknowledgments.tex
Our thanks to Alessandro Acquisti, danah boyd, Miranda Christ, Abraham Flaxman, Terrance Liu, Sarah Radway, Manish Raghavan, and seminar participants at Columbia University, Carnegie Mellon University, and the 2023 NBER Conference on Data Privacy Protection and the Conduct of Applied Research for their feedback. Thanks especially to danah boyd, who first raised this potential privacy issue to our attention, and to Manish Raghavan, whose code inspired the objective in Eq.~\ref{eq:emle} and the swapping algorithm, and was a model for our own codebase. Thanks also to Sarah Young for helping us locate the HUD data used in this study.

%% file: inserts/responsible-disclosure.tex
The attacks described here pose a potentially serious privacy risk to residents living in subsidized housing. The aim of this paper is to evaluate these risks to aid policymakers and public officials in designing future privacy protections. Per Carnegie Mellon University's Institutional Review Board, our research does not meet the criteria for human subjects research and does not qualify for exemption or review, but we are independently taking steps to ensure our research does not help to bring about the potential privacy harms we raise here. While our results suggest that these attacks may be less effective against the new 2020 Decennial Census disclosure avoidance system, people who were living in public housing in 2010 remain potentially vulnerable.

To avoid increasing the risk of harm in the process of disclosing these risks, all of our household-level results are reported using \emph{only synthetic} data. We do not attempt to re-identify any of the households in reconstructions from published data. As an additional precaution, the full results of reconstructions on published data will be kept confidential, we will never attempt to re-identify them, and we report only aggregate statistics. The code and linked data we used to conduct these attacks are only available by request and subject to review, for replication purposes only. We shared earlier versions of this research with both the Census Bureau and Department of Housing and Urban Development, and we encourage them to consider our results when designing privacy protections for current and planned data products.

%% file: inserts/appendix.tex
\renewcommand\thefigure{\thesection.\arabic{figure}}
\setcounter{figure}{0}

\section{Additional Results}

\begin{figure}[!ht]
    \centering
    \includegraphics[width=0.48\linewidth]{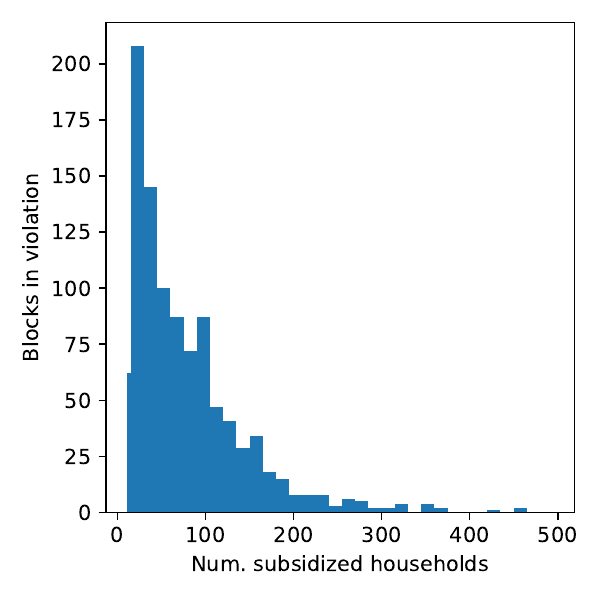}\quad
    \includegraphics[width=0.48\linewidth]{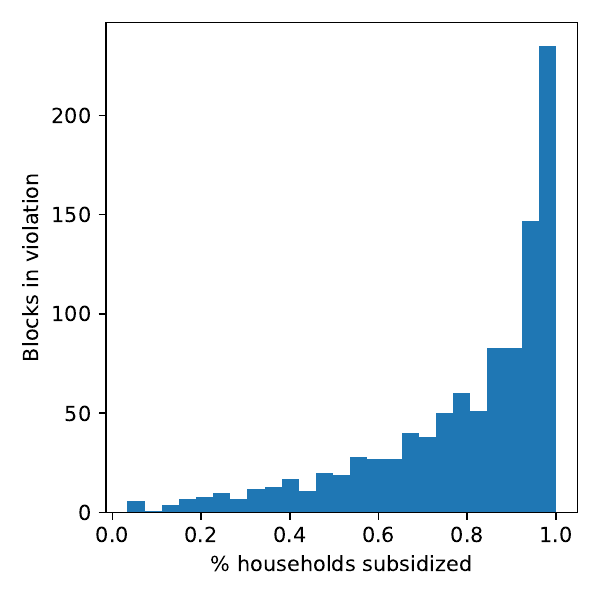}
    \caption{Attack on 2010 public data: number of blocks in violation of occupancy limits by number of subsidized households in the block and by proportion of subsidized households in the block. It may be easier for an attacker to identify which households are living in violation in blocks with fewer subsidized households, but blocks with a high percentage of subsidized households may contain more address matching errors (Appendix~\ref{app:link}).}
    \label{fig:violations-s8reported}
\end{figure}

\begin{figure}[!ht]
    \centering
    \includegraphics[width=0.8\linewidth]{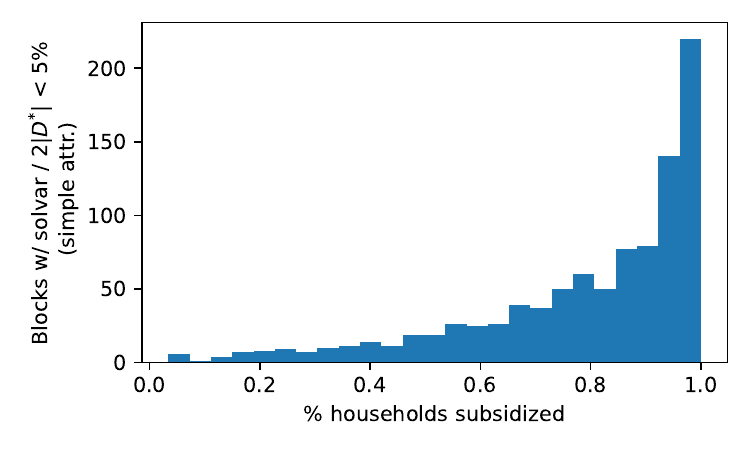}
    \caption{Attack on 2010 public data: distribution of proportion of subsidized households over blocks with solution variability less than 5\% (simplified set of attributes including household size, bedroom count, presence of white, non-Hispanic householder, presence of children).}
    \label{fig:solvar-s8ratio}
\end{figure}

\begin{figure}[!ht]
    \centering
    \includegraphics[width=0.8\linewidth]{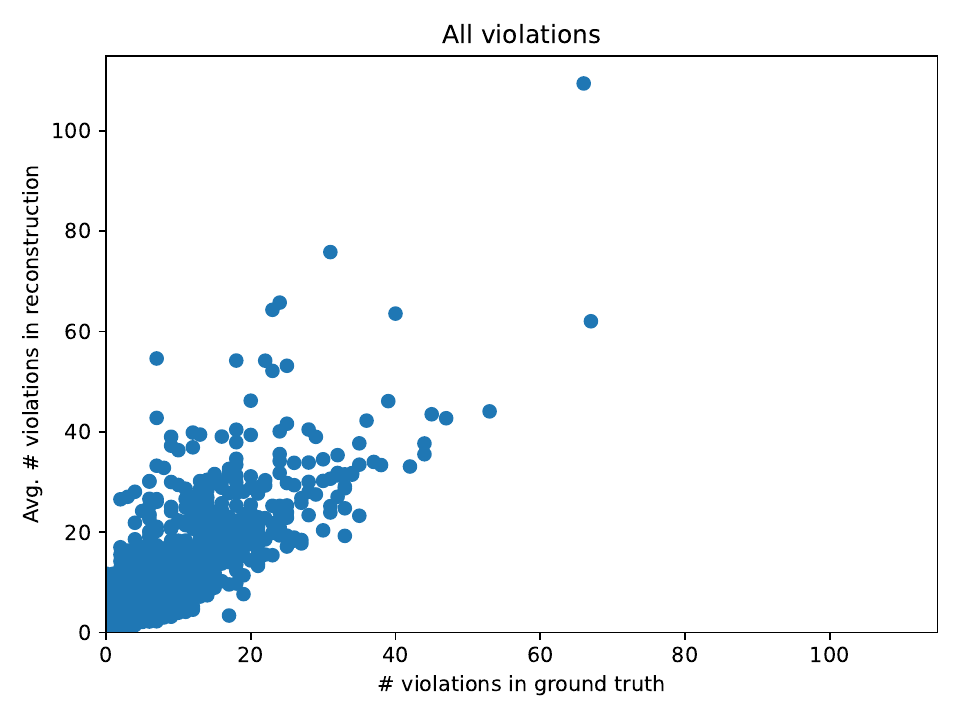}
    \caption{Attack on synthetic data: number of putative violations vs. number of true violations in each block in the swapped, simulated ground truth {\ppmf} microdata.}
    \label{fig:violation-correlation}
\end{figure}

\begin{figure}[!ht]
    \centering
    \begin{subfigure}{0.49\linewidth}
        \centering
        \includegraphics[width=\linewidth]{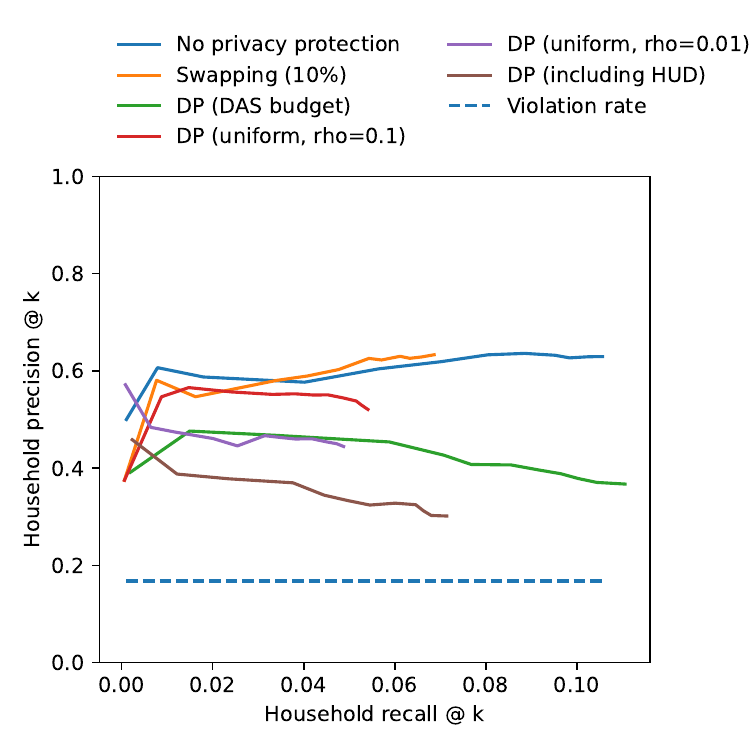}
        \caption{Matching with subsidized status, bedroom counts, householder race/ethnicity, and presence of children (the {\hud} microdata).}
    \end{subfigure}
    \begin{subfigure}{0.49\linewidth}
        \centering
        \includegraphics[width=\linewidth]{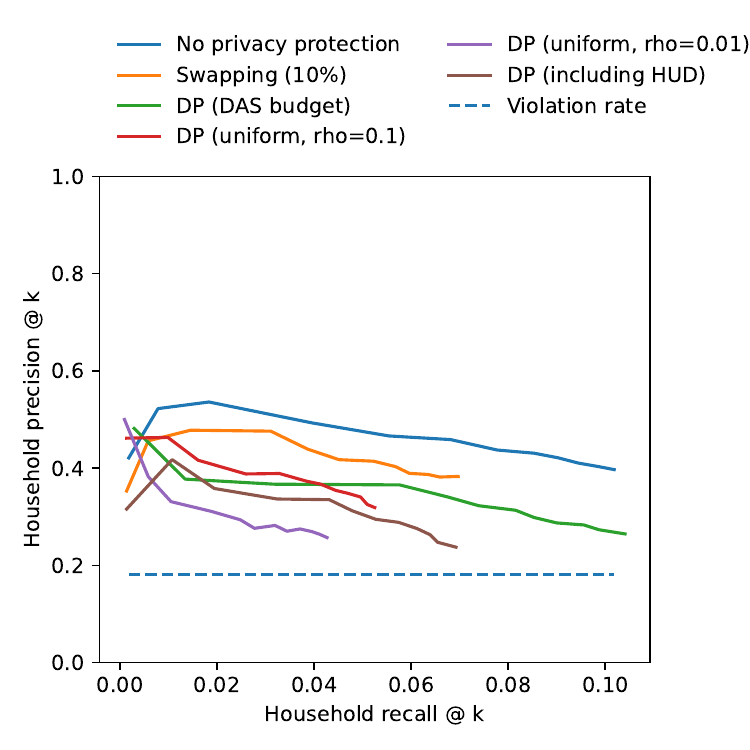}
        \caption{Matching with householder race/ethnicity, and presence of children (no subsidized status or bedroom counts).}
    \end{subfigure}
    \caption{Attack on synthetic data: $\textsc{Precision}_{V,k}(R_\mathcal{S})$ vs. $\textsc{Recall}_{V,k}(R_\mathcal{S})$, when an attacker re-identifies households unique in their block with two different match keys $R_\mathcal{S}$. Swapping at {\synthswaprate} results in fewer putative violations identified and therefore decreases maximum recall compared to the DP mechanisms we tested, but does not significantly reduce precision, recall held equal. (Mechanisms generate different numbers of putative violations depending on the reconstructions produced; maximum recall may be higher more by chance than because the attacker learned more about responding households.) Household-level precision against swapping may even be greater than against no privacy protection because the block-level recall (Eq.~\ref{eq:csp}) also decreases, excluding some blocks that are more difficult to reconstruct.}
    \label{fig:recall-unique}
\end{figure}

\begin{figure}[!ht]
    \centering
    \begin{subfigure}{\linewidth}
        \includegraphics[width=0.49\linewidth]{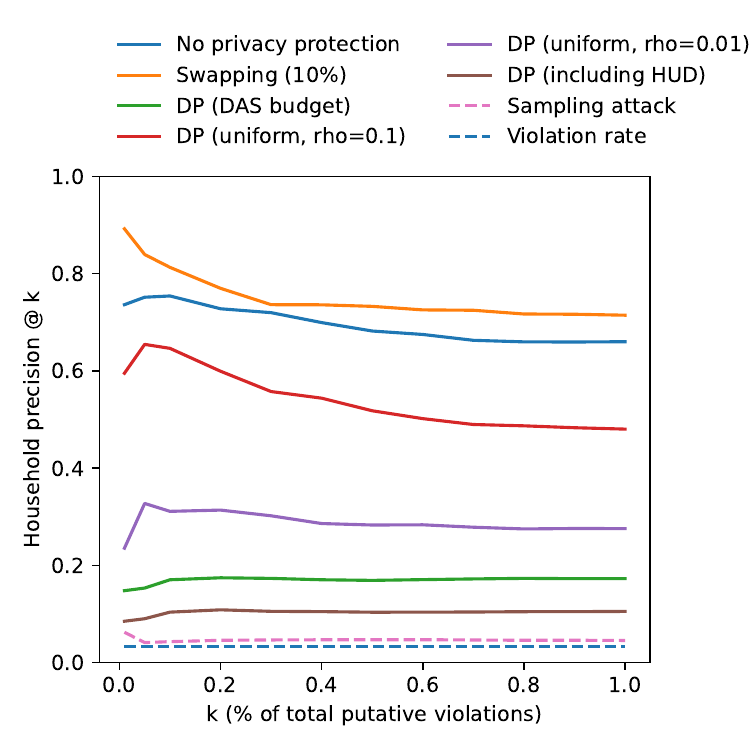}
        \includegraphics[width=0.49\linewidth]{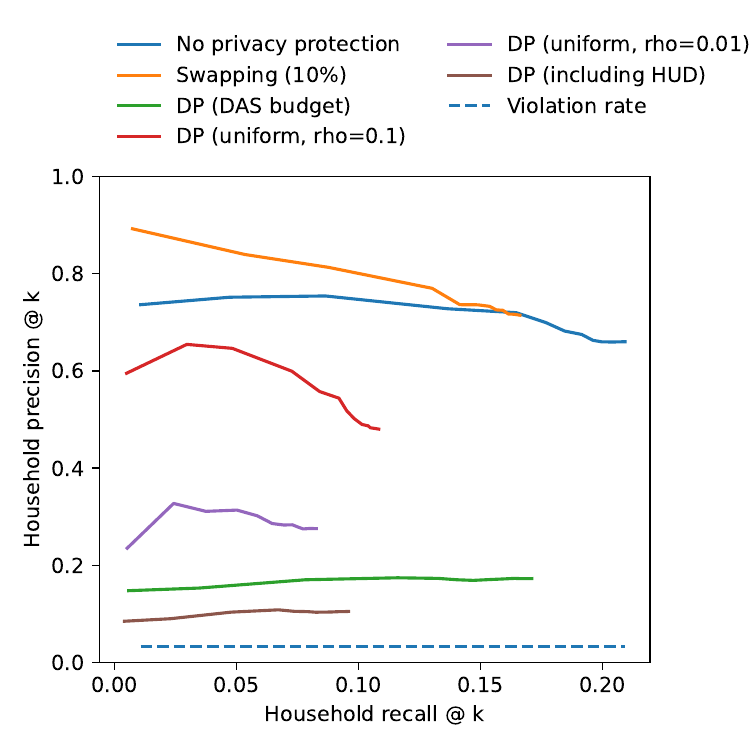}
        \caption{Matching with bedroom counts, householder race/ethnicity, and presence of children (the {\hud} variables).}
        \label{fig:precision-recall-hudsecknown}
    \end{subfigure}
    \begin{subfigure}{\linewidth}
        \includegraphics[width=0.49\linewidth]{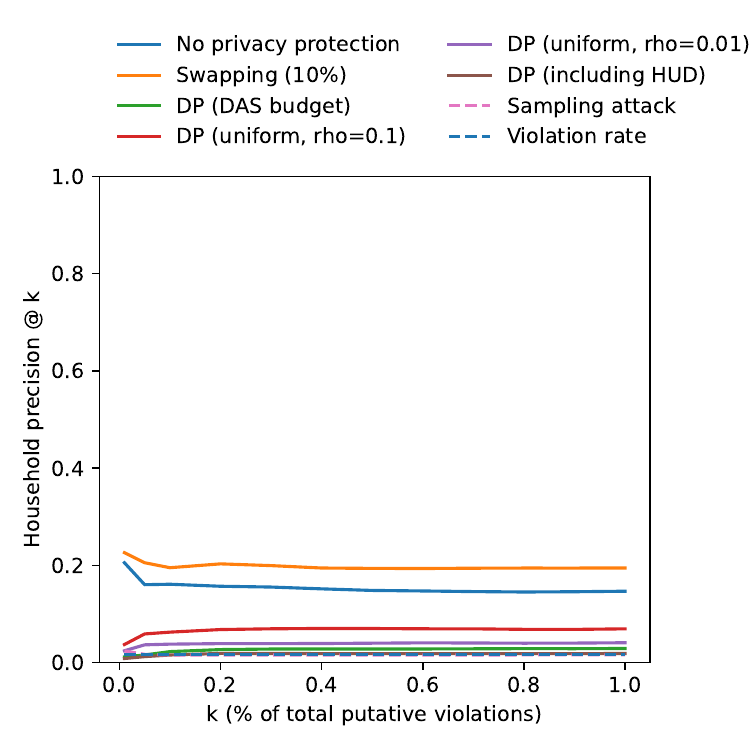}
        \includegraphics[width=0.49\linewidth]{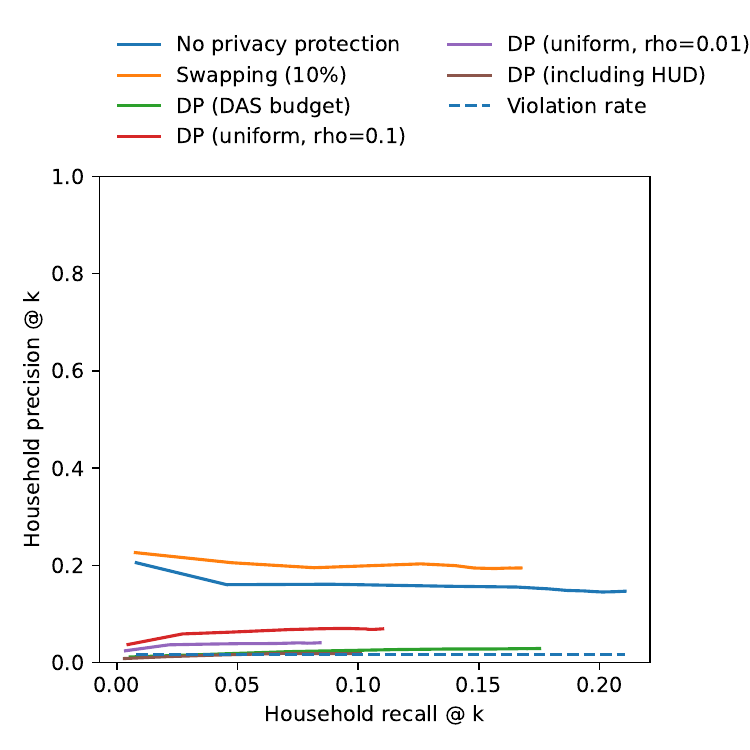}
        \caption{Matching with householder race/ethnicity, and presence of children (no subsidized status or bedroom counts).}
        \label{fig:precision-recall-hhsecunknown}
    \end{subfigure}
    \caption{Attack on synthetic data: Household $\textsc{Precision}_{V,k}(R_\mathcal{S})$ and $\textsc{Recall}_{V,k}(R_\mathcal{S})$, when an attacker re-identifies households with two different match keys $R_\mathcal{S}$. Unlike Figure~\ref{fig:precision-unique}, here we calculate precision and recall over all the subsidized households, not just those unique in their blocks.}
    \label{fig:precision-recall}
\end{figure}

\begin{figure}[!ht]
    \centering
    \includegraphics[width=0.49\linewidth]{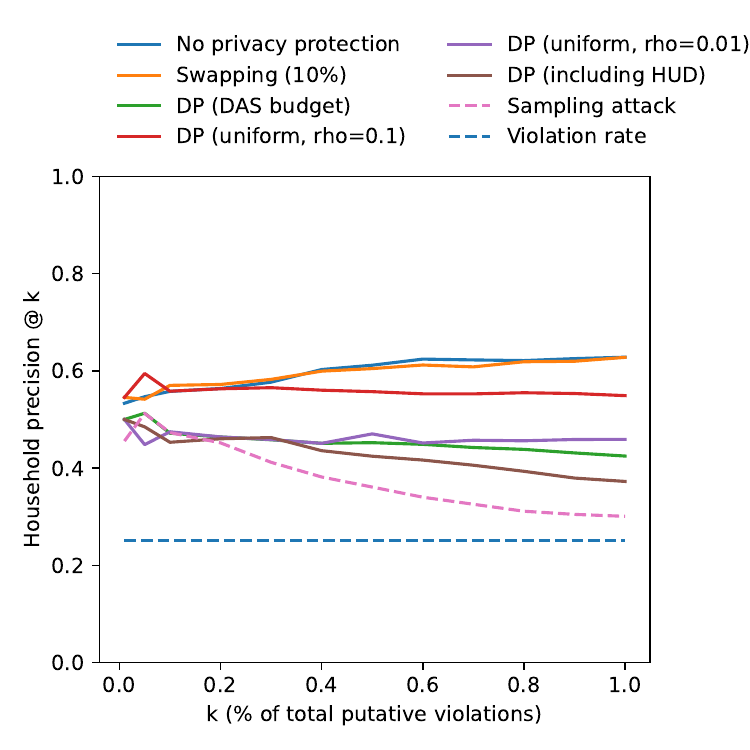}
    \includegraphics[width=0.49\linewidth]{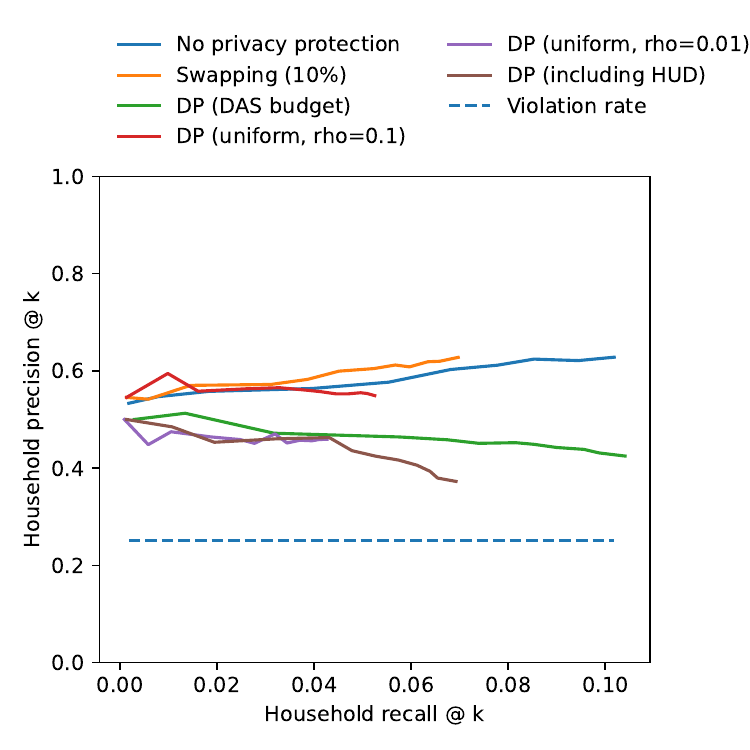}
    \caption{Attack on synthetic data: Household $\textsc{Precision}_{V,k}(R_\mathcal{S})$ and $\textsc{Recall}_{V,k}(R_\mathcal{S})$, when an attacker re-identifies unique households by matching with subsidized status, householder race/ethnicity and presence of children. Unlike Figure~\ref{fig:precision-unique}, here the attacker \emph{does} know which households are subsidized.}
    \label{fig:precision-recall-hh-s8known}
\end{figure}

\clearpage

\section{Synthetic microdata}
\label{app:simulation}
\begin{enumerate}
    \item \textbf{Obtain synthetic {\census} characteristics.} We rely primarily on unit-level privacy-protected microdata files ({\ppmf}), a full set of 2010 demonstration household microdata released with noise from the April vintage of the disclosure avoidance system under production settings \citep{us_census_bureau_just_2023}. While these microdata are not the original, confidential microdata, they are designed to mimic the statistics in the original 2010 {\census}. Because these microdata do not contain exact household sizes over 7, or the exact number of children for households with children, we simulate these values for each household by sampling from their conditional empirical frequency in {\pums} (see below).
    \item \textbf{Simulate {\hud} characteristics.} The synthetic microdata contain only the characteristics summarized in {\census}, so for each block, we must simulate the ``ground truth'' violations---specifically, which households belong to a subsidized property and their bedroom counts. Because we do not have any data on the joint distribution of bedroom count with the {\census} household characteristics, we simulate subsidized status and bedroom counts based on the proportion statistics in {\hud}, assuming those statistics are independent for simplicity (see below).
    \item \textbf{Evaluate queries.} For each block, we run the queries $Q$ on the synthetic microdata $\tilde{D}$ to produce a synthetic version of the summary statistics $Q(\tilde{D})$.\footnote{Note that this validation procedure assumes that the ground truth data for {\census} and {\hud} are the same. In practice, the ground truth might change between collection dates or be recorded differently by the two agencies, which could reduce the effectiveness of reconstruction.}
\end{enumerate}

\inlinesection*{Simulating missing values.} The synthetic microdata from {\ppmf} includes only categorical household sizes up to 7+. So that we can compute statistics on block population, for each record in {\ppmf} with household size >7, we simulate the exact household size by sampling from the empirical distribution of household size in households with matching attributes in {\pums}.
{\ppmf} also does not include the number of children $c$---only a binary indicator for whether a household has children. So similarly, for each record in our synthetic microdata, we sample the number of children in the household from the distribution of number of children in households with matching characteristics in {\pums}.

\inlinesection*{Simulating {\hud} attributes.} {\ppmf} also does not specify which households live in subsidized properties or their bedroom counts---but these variables are necessary for determining whether a synthetic household is living in violation of the ``two heartbeats'' rule. We do not have an empirical estimation of the joint distribution of subsidized status, bedroom counts, and the other {\ppmf} characteristics $\tilde{d}$.
Instead, we take the {\hud} counting queries as a set of independent empirical priors $\{\hat{p}_l\}$ over $\{d_l\}$, the binary attributes counted in {\hud} (householder race/ethnicity $r^j$ and whether the household has children $\ind(c > 0)$). To simulate $\tilde{s}$, a binary indicator for whether a household $\tilde{h}$ lives in a subsidized property, we draw $N_s$ households for each block with the heuristic likelihood
\begin{align}
    \frac{\prod p_l^{\tilde{d}_l}(1-p_l)^{1-\tilde{d}_l}}{1/M \sum_{k=1}^M \ind(d_k = \tilde{d})}.
\end{align}
Similarly, to simulate the number of bedrooms in each subsidized household, we take the empirical probability $\hat{p}_b$ of a bedroom count $b$ (0 or 1, 2, or 3+) as an independent prior. Then, we control the violation rate with a parameter $\alpha$, sampling $\tilde{b}$ from $\hat{p}_b\alpha^{\text{violation}(b, \tilde{x})}$. Our main results set $\alpha=\synthalpha$ so as to reduce the synthetic rate of violation among subsidized households to {\synthpercentviolationsec}, close to the rate reconstructed from public data. Violations are spread fairly evenly across blocks in our simulation---{\synthblocksviolationpercent} of blocks have at least one violation.

\section{Adapting to noisy queries} 
\label{app:soft}
Solving for the exact $Q(D)=Q(D')$ is sometimes impossible when noise is added to the queries. (For example, it is impossible to find a feasible reconstruction for {\dpipinvalidpercent} of blocks in the differentially private demonstration statistics.) To make this attack feasible for noisy queries, we could instead minimize the query error \citep{dick_confidence-ranked_2022}, where $L(D')$ is either $0$ for noisy constraint satisfaction or the empirical likelihood function from Eq.~\ref{eq:emle}:
\begin{align}
    \argmin_{D'} -\log L(D') + \| Q(D') - Q(D) \|_2^2 \label{eq:soft}
\end{align}

\section{Linking the 2010 Census to HUD Statistics}
\label{app:link}
We focus on properties receiving assistance from the following housing programs: Section 8 New Construction and Substantial Rehabilitation Program (Section 8 NC/SR), Section 236 Projects, public housing, and multifamily assisted projects with Federal Housing Administration insurance or HUD subsidies. On average, {\hudresponserate} of households in each subsidized property are recorded in the HUD data---we conservatively assume that these are the only subsidized households in the block, and that they live in the largest units reported in {\hud}.

Since our HUD data did not contain information about which census block each property belongs to besides the property address, we identified the census block for each property using the Census Geocoder \citep{us_census_bureau_census_2023}. We filtered out {\numpropertiesnogeocoderresult} properties where the Census Geocoder was unable to find a county, tract code, and block code (for example, because the property spanned multiple addresses, or the address was invalid). There were also {\numblockswithsuproperties} blocks matched with more than one subsidized property address---for these blocks, we created a single partition of subsidized households by aggregating the summary statistics across properties within a block.

Because our attack operates on household sizes, we also removed {\numblockswithGQ} blocks where a portion of the population lived in group quarters (e.g. colleges, correctional facilities) for simplicity---however, our approach could be modified to accommodate group quarters. Of these remaining blocks, we then ignored {\numblocksoccupiedhhexceed} blocks where the number of households in the block provided by Summary File 1 was less than the number of occupied subsidized units reported by HUD as well as {\numblockspeopletotalexceed} blocks where the number of people reported living in subsidized properties was greater than the number of people reported living in the block in Summary File 1. These discrepancies likely occur when the address of the property in our HUD dataset is listed in a different block than the households themselves---for example, when the leasing company’s building is located in a different block from the housing property---or the property may exceed the bounds of a single block, especially for properties with many subsidized households. We leave out these blocks and assume for simplicity that the rest are properly addressed, although there may be additional address matching errors we did not detect.

There are also {\numblocksbrrounding} blocks where the percentage of 0 or 1 bedroom, 2 bedroom, and 3 bedroom units could not be precisely rounded to a set of integers summing to the number of reporting subsidized households. To be as conservative as possible, for blocks where the number of households is greater than the number of units, we assume the $N_\text{unassigned}$ largest subsidized households are not in violation, where $N_\text{unassigned}$ denotes the difference between the number of subsidized households and the number of subsidized units. Typically, $N_\text{unassigned} \leq 1$.

\section{Integer Programming}
\label{app:ip}
For a given block, suppose there are $N$ households, and that households $1, 2, \ldots, N_s$ are located in a subsidized property. For each household $i$, an attacker aims to reconstruct the following variables:
\begin{itemize}
    \item $p_i \in \{0, 1, \ldots,  $\text{SF1}[P]$\}$: the number of people in a given household $i$, which we conservatively assume can be no more than $\text{SF1}[P]$, the total block population recorded in Summary File 1.
    \item $r^j_i \in \{0, 1\}$: a binary indicator of whether household $i$ contains a person of race/ethnicity $j \in J$ (e.g. White and non-Hispanic, White and Hispanic, etc.)
    \item $c_i \in \{0, 1, \ldots, p_i\}$: the number of people under 18 in household $i$.
    \item $b^k_i \in \{0, 1\}$: a binary indicator of whether the household lives in a unit with $k \in \{\leq 1, =2, \geq 3\}$ bedrooms.
\end{itemize}
We select the following summary statistics from Summary File 1 from the 2010 Decennial Census (SF1) and the 2010 HUD data (HUD):\footnote{Specifically, we used data from these tables from {\census}: (1) P1 Total Population; (2) P5 Hispanic or Latino Origin by Race; (3) P15 Hispanic or Latino Origin of Householder by Race of Householder; (4) P16 Population in Households by Age; (5) H3 Occupancy Status; (6) H13 Household Size.}
\begin{itemize}
    \item $\text{SF1}[N_{p=x}]\;\forall\; x \in \{1, 2, \ldots, 6\}$: the number of households of size $x$ reported in Summary File 1.
    \item $\text{SF1}[P]$: the block population reported in table P1 in Summary File 1.
    \item $\text{HUD}[P]$: the total number of residents reported in the HUD data.
    \item $\text{SF1}[N_{j}]  \; \forall \;  j \in J $: the number of households with a person of race/ethnicity $j$ reported in Summary File 1.
    \item $\text{HUD[Householder of race $j$]} \; \forall \;  j \in J_{-h}$: the number of households with a householder of non-Hispanic race $j$ reported in the HUD data.
    \item $\text{HUD[Hispanic householder]}$: the number of households with a Hispanic householder reported in the HUD data.
    \item $\text{SF1}[N_{c}]$: the number of people under 18 (children) reported in Summary File 1.
    \item $\text{HUD[1 adult with children]}$, $ \text{HUD[2 adults with children]}$: the number of households with 1 adult with children or 2 adults with children, respectively, reported in the HUD data.
    \item $\text{HUD[Units with $k$ bedrooms]} \; \forall \;  k \in \{\leq 1, =2, \geq 3\}$: the number of households with $x$ bedrooms reported in the HUD data.
\end{itemize}
We constrain the entire block with Summary File 1 statistics:
\begin{itemize}
    \item $\left(\sum^N_{i=1} \ind(p = x) \geq \text{SF1}[N_{p_i=x}]\right) \; \forall \;  x \in \{1, 2, \ldots, 6\}$: the total number of households of size $x$ matches Summary File 1. (In some cases, the total number of households is greater than the sum of the household size distribution, usually by a very small margin. We allow the remaining household(s) to take any size.)
    \item $\sum^N_{i=1} p_i \geq \text{SF1}[P]$: the total number of people in the block is at least the block population reported in Summary File 1. (We use a lower bound for this and similar statistics in case of nonresponse.)
    \item $\left(\sum^N_{i=1} r^j_i \geq \text{SF1}[N_{j}]\right) \; \forall \;  j \in J$: the total number of households with a person of race/ethnicity $j$ is at least the number of households with a person of race/ethnicity $j$ reported in Summary File 1.
    \item $\sum^N_{i=1} c_i \geq \text{SF1}[N_{c}]$: the total number people under 18 matches Summary File 1.
\end{itemize}
We also constrain the first $N_s$ households by the statistics reported in the HUD data:
\begin{itemize}
    \item $\sum^{N_s}_{i=1} p_i \geq \text{HUD}[P]$: the total number of people in subsidized properties in the block is at least a) the total number of residents reported in the HUD data or b) the total size of the largest possible set of $N_s$ subsidized households from {\census}, whichever is smaller. (In a few cases of discrepancy between the two datasets, this number is not the same; in those cases, we conservatively choose the smaller lower bound.)
    \item $\left(\sum^{N_s}_{i=1} r^j_i \geq \text{HUD[Householder of race $j$]}\right) \; \forall \;  j \in J_{-h}$: the total number of subsidized households with any non-Hispanic person of race $j \in J_{-h}$ is at least the number of households with a person of race $j$ reported in the HUD data. 
    \item $\sum^{N_s}_{i=1} \prod^{J_h}_{j=1} r^j_i \geq \text{HUD[Hispanic householder]}$: the total number of subsidized households with any person of Hispanic origin is at least the number of households with a Hispanic householder reported in the HUD data. (As above, {\hud} only counts the householder.)
    \item $\sum^{N_s}_{i=1} \ind(c_i > 0) = \text{HUD[1 adult with children]} + \text{HUD[2 adults with children]}$: the total number of subsidized households with people under 18 matches the total number of households with children reported in the HUD data.
    \item $\left(\sum^{N_s}_{i=1} b^k_i \geq \text{HUD[Units with $k$ bedrooms]}\right) \; \forall \;  k \in \{\leq 1, =2, \geq 3\}$: the number of subsidized households with $x$ bedrooms is at least the number of households with $x$ bedrooms reported in the HUD data. In some cases, not every subsidized unit in {\hud} reports a bedroom count, so we treat the bedroom counts as minimums, allowing the solver to add any number of bedrooms to the units with missing values.
\end{itemize}
Finally, to detect blocks with at least one household in violation, we constrain the bedroom size in subsidized units according to the ``two heartbeats per room'' rule: $\sum^{N_s}_{i=1} b^{\leq 1}_i \ind(p_i > 2) = 0$ and $\sum^{N_s}_{i=1} b^{=2}_i \ind(p_i > 4) = 0$. If a reconstruction satisfying all these constraints does not exist, the block must contain a household in violation of this occupancy limit.

\section{DAS Replication}
\label{app:budget}
We simplify the 2020 disclosure avoidance system \citep{abowd_2020_2022} in two ways:
\begin{enumerate}
    \item The TopDown algorithm used for the 2020 Census adaptively allocates privacy budget to many queries over multiple geographic levels with the matrix mechanism \citep{li_matrix_2015}, a two-stage process that first applies noise to a set of ``strategy'' queries, then combines those strategy queries to produce the full set of published queries. As a result, there is no clear privacy budget allocation to any of the non-strategy queries (e.g., household size) used in our attack. To replicate the privacy budget allocated to each non-strategy query $q$ used in our attack, we calculate the total variance applied to $q$ in each strategy query in the DAS budget. We then add noise to $q$ using the lowest variance across strategy queries (usually the strategy query with the fewest cross products or the highest budget; the exact per-query allocations we used are below).
    \item While the disclosure avoidance system includes multipass optimization post-processing to create consistency within tables, we enforce consistency with a simple bottom-up optimization. Like the TopDown algorithm, we also treat the number of occupied units as invariant \citep{abowd_2020_2022}; we re-scale all noisy queries from the household tables to match this value. Also, after our post-processing, the total population in the housing tables is consistent with the total population in the person tables.
\end{enumerate}

The Census Bureau publishes the privacy budget as a set of allocations to strategy queries sufficient to construct the person and household tables \citep{abowd_2020_2022, us_census_bureau_privacy-loss_2023}. There is no direct mapping between the budget and an arbitrary query $q$. So, we replicate the budget applied to each query $q$ (e.g., the number of households of a given size) following guidance from census officials:
\begin{enumerate}
    \item Identify all variables $x$ in the strategy queries that are sufficient to answer the query $q$. (E.g., household size appears in \texttt{DETAILEDCOUPLETYPEMULTGENDETOWNCHILDSIZE} and \texttt{DETAILED}.) The definitions of these queries can be found in the 2020 DAS Github repository.\footnote{\url{https://github.com/uscensusbureau/DAS_2020_DHC_Production_Code}} Determine the number of subqueries (cells in the cross-tabulation) needed to sum to the query $q$ on the margin (e.g., in \texttt{DETAILEDCOUPLETYPE...}, \texttt{HH\_SIZE} crosses with \texttt{COUPLE\_TYPE} (5 values), \texttt{MULTIG} (2 values), and \texttt{CHILD} (4 values) for a total of $5\times 2 \times 4 = 40$ cells).
    \item Then, identify all the strategy queries containing one or more intermediate queries $x$. (E.g. \texttt{DETAILEDCOUPLETYPE...} appears in both
    \texttt{SEX * HISP * HHTENSHORT\_3LEV * RACE * DETAILEDCOUPLETYPE...} and one other query strategy.) For each strategy query identified, \begin{enumerate}
        \item Determine the number of subqueries (crosstab cells) required to compute $x$ on the margin, and use this number to determine the number of subqueries $M_{qi}$ required to compute $q$ from $i$. (E.g., \texttt{SEX} has 2 values, \texttt{HISP} 2, \texttt{HHTENSHORT\_3LEV} 3, and \texttt{RACE} 7. Each of the 40 cells needed to compute \texttt{HH\_SIZE} from \texttt{DETAILEDCOUPLETYPE...} is crossed with these variables in the strategy query above. So the total number of cells needed to sum to any of the 7 \texttt{HH\_SIZE} attributes is $M_\texttt{HH\_SIZE, SEX * HISP * \ldots} = 40 \times 2 \times 2 \times 3 \times 7 = 3,360$.
        \item Each query strategy $i$ receives $c_i\rho$ privacy budget, where $\rho$ is the global budget for the table. (E.g. for the query strategy above, $c_\texttt{SEX * HISP * \ldots} = 0.0002$. DAS adds Gaussian noise $\sigma_{ik}^2 = \frac 1 {c_i\rho}$ to each of the crosstabs comprising each strategy query $i$ \citep{abowd_2020_2022, canonne_discrete_2022}. So the total variance to compute the query $q$ is $\sum^{M_{qi}}_{k=1} \sigma_{ik}^2 = \frac {M_{qi}} {c_i\rho}$, equivalent to budget $\rho c_i / M_{qi}$. (E.g., \texttt{HH\_SIZE} receives a budget of $\rho 0.0002/3360$ when computed using the query strategy above.)
    \end{enumerate}
    \item Use the highest adjusted budget allocation $\rho_{iq} = c_i / M_{qi}$ across all the strategy queries $\{i\}$ that are sufficient to compute $q$.
\end{enumerate}

This strategy results in the following per query budget allocations for statistics from our simulated, differentially private version of SF1, as proportions of the global budget $\rho$. Using the notation from Appendix~\ref{app:ip}:
\begin{itemize}
    \item $\text{SF1}[N_{p_i=x}]$, the number of households of size $x \in \{1, \ldots, 7+\}$: $0.0002 / 3360$
    \item $\text{SF1}[N_{j}]  \; \forall \;  j \in J $, the number of households with a person of race/ethnicity $j$: $0.0002 / 2$
    \item $\text{SF1}[N_{c}]$, the number of children: $0.0002 / 2$
\end{itemize}
For our reconstruction attack, it is sufficient to set the block population $\text{SF1}[P]$ equal to the sum of the household distribution, letting all households size $7+$ be size $7$.

While the disclosure avoidance system for the 2020 Decennial Census uses complex multipass optimization to reconcile noisy statistics into a consistent set of tables, our post-processing is simpler. First, like the Census Bureau, we clip all negative counts to zero. Second, because we follow the TopDown algorithm \citep{abowd_2020_2022} in treating the number of occupied households as invariant, we rescale the noisy household size queries $\text{SF1}[N_{p_i=x}]$ and the household race/ethnicity queries $\text{SF1}[N_{j}]  \; \forall \;  j \in J $ such that their respective sums exactly equal the number of occupied units. Second, statistics that are aggregations of other statistics are computed directly, bottom-up. (For example, the total number of households with Hispanic members can be computed from the household race/ethnicity queries.) Third, we upper bound the number of children by the total population, and we set the total population to the sum of the household size distribution.

\section{Match Rate}
\begin{figure}[!ht]
    \centering
    \begin{subfigure}{0.49\linewidth}
        \centering
        \includegraphics[width=\linewidth]{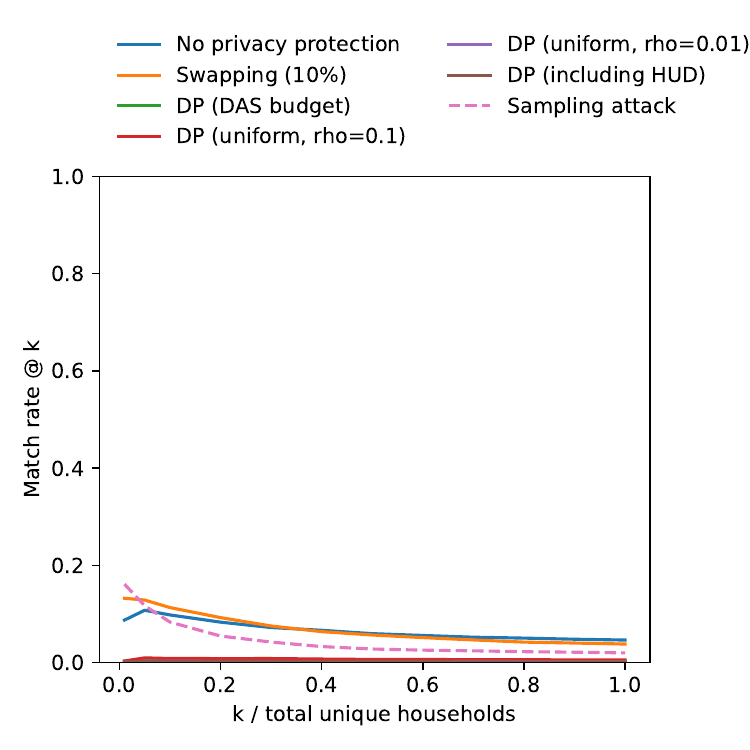}
        \caption{Matching on all variables.}
    \end{subfigure}
    \begin{subfigure}{0.49\linewidth}
        \centering
        \includegraphics[width=\linewidth]{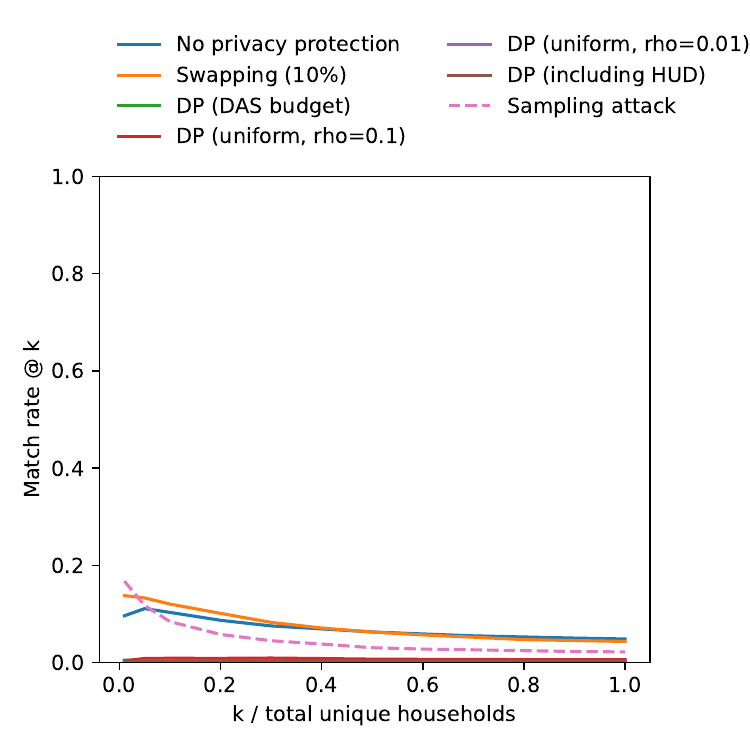}
        \caption{Matching on just {\census} variables.}
    \end{subfigure}
    \begin{subfigure}{0.49\linewidth}
        \centering
        \includegraphics[width=\linewidth]{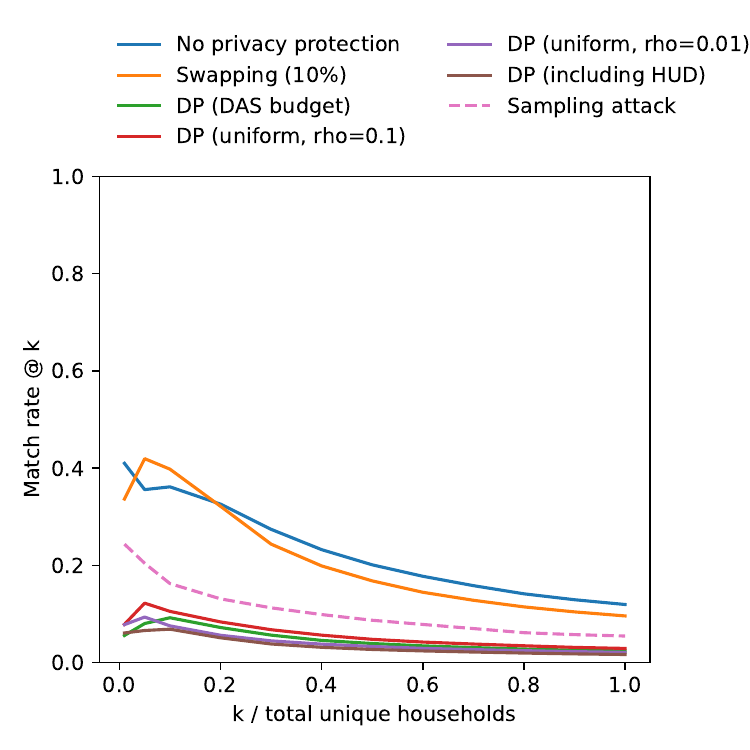}
        \caption{Matching on just {\hud} variables.}
    \end{subfigure}
    \begin{subfigure}{0.49\linewidth}
        \centering
        \includegraphics[width=\linewidth]{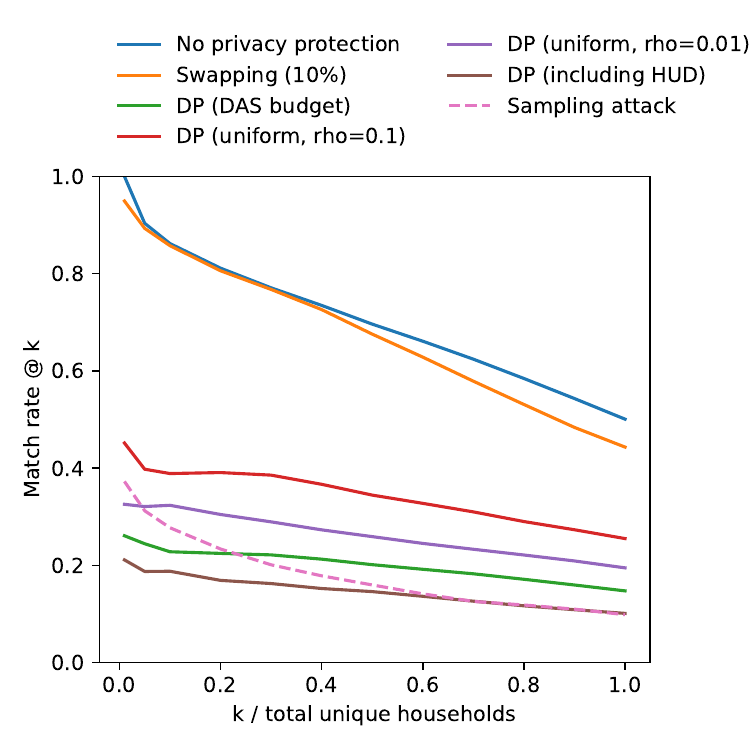}
        \caption{Matching on household size and bedrooms.}
    \end{subfigure}
    \caption{$\textsc{Match Rate}_{V,k}(R_\mathcal{S})$, matching putative violations with violating households using various match keys $R_\mathcal{S}$. As in Figure~\ref{fig:precision-unique}, we compute the match rate only for the putative violations from Eq.~\ref{eq:csp}.}
    \label{fig:match-rate}
\end{figure}

Our main results focus on precision and recall---the likelihood that an attacker would have success matching putative violations to real households or individuals. Another common way to measure the quality of ranked reconstructions is by the number of reconstructed records that actually appear in the data \citep{dick_confidence-ranked_2022}. We narrow this metric to focus only on the number of most frequent reconstructed putative violations that match at least one violating household in our simulated ground truth.
\begin{align}
    \textsc{Match Rate}_{V,k}(R_\mathcal{S}) \leftarrow \frac{1}{k} \sum_{i=1}^k \ind\{R_i \in V\}
\end{align}
The violation match rate for our reconstructions is displayed in Figure~\ref{fig:match-rate}, and tends to correspond qualitatively with the precision scores. The match rates are relatively low---many reconstructed households in violation do not actually appear in the data. But, the attacker can ignore these households when they do not correspond to a real household in the match key $R_{\mathcal{S}}$.

